\documentclass{aa}
\usepackage[varg]{txfonts}

\usepackage{natbib}
\bibpunct{(}{)}{;}{a}{}{,} 
\usepackage{graphicx}	
\usepackage{amsmath}	

\usepackage[hyperfootnotes=false, linktocpage=true, breaklinks=true, colorlinks=true, linkcolor=blue, citecolor=blue, urlcolor=blue]{hyperref}

\usepackage{placeins}

\usepackage{longtable}
\usepackage{xcolor}

\usepackage{booktabs}

\newcommand{\funit}{~erg cm$^{-2}$ s$^{-1}$} 
\newcommand{\ergs}{~erg s$^{-1}$}   
\newcommand{\erg}{~erg}  
\newcommand{\cmcube}{~cm$^{-3}$}

\newcommand{\chisq}{$\rm{\chi^{2}}$ }
\newcommand{\chisqr}{$\rm{\chi_{\nu}^{2}}$ }

\newcommand{\msun}{\rm M_\odot} 
\newcommand{\pc}{\,\rm pc} 
\newcommand{\kpc}{\,\rm kpc} 

\def\xmm{{\it XMM-Newton}} 
\def\mass{{\it 2MASS}}

\def\chandra{{\it Chandra}}
\def\suzaku{{\it Suzaku}}
\def\ginga{{\it Ginga}}
\def\nustar{{\it NuSTAR}}

\def\fermi{{\it Fermi}} 
\def\erosita{{\it eROSITA}} 
 
\def\meerkat{{\it MeerKAT}}

\def\spitzer{{\it Spitzer}}
\def\irac{{\it IRAC}}

\begin{document}

\title{Study of the excess Fe~XXV line emission in the central degrees of the Galactic centre using \xmm{} data}

\author{K. Anastasopoulou\inst{1}
\and G. Ponti\inst{1,2}
\and M. C. Sormani\inst{3}
\and N. Locatelli\inst{2} 
\and F. Haberl\inst{2}
\and M. R. Morris\inst{4} 
\and E. M. Churazov\inst{5,6}
\and R. Schödel\inst{7}
\and C. Maitra\inst{2}
\and S. Campana\inst{1}
\and E. M. Di Teodoro\inst{8}
\and C. Jin\inst{9,10}
\and I. Khabibullin\inst{11,5,6}
\and S. Mondal\inst{1}
\and M. Sasaki\inst{12}
\and Y. Zhang\inst{2}
\and X. Zheng\inst{2}}

\institute{INAF-Osservatorio Astronomico di Brera, Via E. Bianchi 46, I-23807 Merate (LC), Italy\\ \email{konstantina.anastasopoulou@inaf.it}
\and  Max-Planck-Institut für extraterrestrische Physik, Gie{\ss}enbachstra{\ss}e 1, D-85748, Garching, Germany 
\and  Universität Heidelberg, Zentrum für Astronomie, Institut für theoretische Astrophysik, Albert-Ueberle-Str. 2, 69120 Heidelberg, Germany
\and  Department of Physics and Astronomy, University of California, Los Angeles, CA 90095-1547, USA 
\and  Max Planck Institute for Astrophysics, Karl-Schwarzschild-Str. 1, D-85741 Garching, Germany
\and   Space Research Institute (IKI), Profsoyuznaya 84/32, Moscow 117997, Russia
\and  Instituto de Astrofísica de Andalucía (IAA-CSIC), Glorieta de la Astronomía S/N, 18008, Granada, Spain
\and  Department of Physics and Astronomy, University of Florence, 50019 Sesto Fiorentino, Italy 
\and  National Astronomical Observatories, Chinese Academy of Sciences, 20A Datun Road, Beĳing 100101, China
\and School of Astronomy and Space Sciences, University of Chinese Academy of Sciences, 19A Yuquan Road, Beĳing 100049, China
\and Universitäts-Sternwarte, Fakultät für Physik,
Ludwig-Maximilians-Universität München, Scheinerstr.1, 81679 München,
Germany 
\and  Dr. Karl Remeis Observatory, Erlangen Centre for Astroparticle Physics, Friedrich-Alexander-Universität Erlangen-Nürnberg,
Sternwartstraße 7, 96049 Bamberg, Germany
}

\date{Received xxxx / Accepted xxxx}

\abstract{The diffuse Fe~XXV (6.7\,keV) line emission observed in the Galactic ridge is widely accepted to be produced by a superposition of a large number of unresolved X-ray point sources. In the very central degrees of our Galaxy, however, the existence of an extremely hot ($\sim$7\,keV) diffuse plasma is still under debate. In this work we measure the Fe~XXV line emission using all available \xmm{} observations of the Galactic centre (GC) and inner disc ($-10^{\circ}<\ell<10^{\circ}$, $-2^{\circ}<b<2^{\circ}$). 
We use recent stellar mass distribution models to estimate the amount of X-ray emission originating from unresolved point sources, and find that within a region of $\ell=\pm1^{\circ}$ and $b=\pm0.25^\circ$ the 6.7\,keV emission is 1.3 to 1.5 times in excess of what is expected from unresolved point sources. The excess emission is enhanced towards regions where known supernova remnants are located, suggesting that at least a part of this emission is due to genuine diffuse very hot plasma.
If the entire excess is due to very hot plasma, an energy injection rate of at least  $\sim6\times10^{40}$ \ergs{} is required, which cannot be provided by the measured supernova explosion rate or past Sgr\,A$^{*}$ activity alone. However, we find that almost the entire excess we observe can be explained by assuming GC stellar populations with iron abundances $\sim$1.9 times higher than those in the bar/bulge, a value that can be reproduced by fitting diffuse X-ray spectra from the corresponding regions. Even in this case, a leftover X-ray excess is concentrated within $\ell=\pm0.3^{\circ}$ and $b=\pm0.15^\circ$, corresponding to a thermal energy of $\sim2\times10^{52}$ erg, which can be reproduced by the estimated supernova explosion rate in the GC. Finally we discuss a possible connection to the observed GC \textit{Fermi}-LAT excess.}

\keywords{Galaxy: bulge -- Galaxy: centre -- Galaxy: disc -- X-rays: general -- X-rays: ISM}
\authorrunning{K. Anastasopoulou et al.}
\titlerunning{Fe~XXV line emission in the GC} 
\maketitle



\section{Introduction}\label{introduction}

Studies of galaxies in X-rays have revealed that diffuse X-ray emission is a dominant component of the total soft X-ray flux \citep[$<$2 keV; e.g.][]{mineo12}. It is believed to be the result of the energy released in the interstellar medium (ISM) by supernova explosions and stellar winds \citep{cox74,spitzer90, mckee95}. This feedback on the ISM, in the most extreme cases, takes the form of a galactic wind \citep[e.g. M82;][]{strickland09}, which can drive powerful outflows to the halo.

Studies at high angular resolution, that is to say in our own Galaxy, have recently revealed the detection of soft X-ray emitting bubbles, the \erosita\ bubbles \citep{predehl20}, that extend approximately 14\,kpc above and below the Galactic centre (GC). These features are double the size and about ten times the volume of the \fermi\ bubbles \citep{su10}, and they are most likely the result of large energy injections from the GC. 
Indeed, looking at the central hundred parsecs of the GC,  \citet{ponti19} discovered the Chimneys, which are two prominent X-ray features that extend hundreds of parsecs above and below the Galactic plane, and \citet{heywood19}, studying \meerkat{} radio maps, found two edge-brightened lobes that approximately trace the edges of the X-ray chimneys. These features, along with  the \fermi\ and \erosita\ bubbles, are most likely tracers of an outflow. The driving mechanism of this putative outflow could either be attributed to past activity of Sgr\,A$^{*}$ or to star formation via the production of core-collapse supernovae \citep[][]{zhang21} that possibly result in the existence of a very hot ($\sim$7\,keV) unbound plasma component in the centre of our Galaxy \citep[e.g.][]{koyama96}.

In the late 1970s, the Galactic ridge X-ray emission was discovered as a large, diffuse feature along the Galactic disc and bulge extending about 100$^{\circ}$ along the Galactic plane \citep{cooke69,worrall82}. It showed strong emission lines and a hard X-ray continuum characteristic of a 5-10 keV optically thin thermal plasma \citep{koyama86,yamauchi93,koyama07,yamauchi09}. However, such a hot plasma could not be gravitationally or magnetically bound to the Galaxy \citep{tanaka02}, and it would flow away with supersonic velocity of a few thousand $\rm km\, s^{-1}$ on a timescale of $\sim$3$\times10^{4}$\,yr \citep[][]{zeldovich66,sunyaev93}. Moreover, in order for the hard diffuse emission to be attributed to hot plasma, a steady source of $10^{43}$ \ergs{} would be required to sustain the plasma, and no evidence of such a source exists \citep[see review from][]{tanaka02}.

Alternatively this hard diffuse emission could be composed of a large population of weak point sources \citep{worrall82,worrall83,koyama86,ottmann92,mukai93}. In support of this, it was observed 
that the hard X-ray emission and the Fe~XXV 6.7\,keV line are very well correlated over the whole Galaxy (GC and plane) with the near-infrared (NIR) luminosity (3-4\,$\mu m$), which traces the stellar mass density \citep{revnivtsev06a,revnivtsev06b}. Moreover, with an ultra-deep \chandra{} observation of a field at ($\ell$, $b$) = (0.113$^\circ$, -1.424$^\circ$) on the Galactic plane, \citet{revnivtsev09} resolved more than 80\% of the diffuse 6-8 keV emission into weak discrete sources such as accreting white dwarfs and coronally active stars. The remaining 10-20\% of the total diffuse emission was attributed to stars of luminosities lower than the detection limit of \chandra\ (0.5-7.0 keV; $\lesssim10^{29}$\ergs).

Today all studies agree that $\sim$70-80\% of the Galactic ridge X-ray emission flux is resolved into point sources \citep[see review by][]{koyama18} consisting of some mixture of magnetic and non-magnetic cataclysmic variables (mCVs, non-mCVs), and coronally active sources such as active binaries (ABs) \citep[e.g.][]{revnivtsev09,hong12,morihana22,schmitt22}. However, in the very central degrees of our Galaxy, the existence of a very hot, extended interstellar plasma is still under debate. 

Closer to the GC, with a total of $\sim$600 ks of \chandra{} observations, 20-30\% of the total X-ray flux was resolved into point sources in a sub-region free from supernova remnants (SNRs) \citep[region `Close';][]{muno04, park04}. Moreover, \citet{park04} discussed that the remaining emission could be due to magnetically confined truly diffuse hot plasma.
In addition, \citet{revnivtsev07}, using even deeper \chandra{} observations (918 ks) towards a region south-west of the GC (2-4 arcmin from Sgr\,A$^{*}$) where the contribution of SNRs to the thermal emission is small, found that at least $\sim$40\% of the total X-ray emission in the energy band 4-8 keV originates from point sources ($\mathrm{L_{2-10\, keV} }>$ $10^{31}$\ergs). They found that most of the unresolved X-ray flux possibly originates from weak CVs and coronally active stars with luminosities below the \chandra{} detection threshold. However, they note that the GC region is characterised by an increased number density of SNRs with respect to the Galactic disc, so a small contribution from a truly diffuse emission component can be expected.

A bright peak of 6.7\,keV iron line emission in the GC was first  discovered by the \ginga\ satellite \citep{koyama89}. Subsequently, \citet{yamauchi93} made a map of the same emission and showed that it has a roughly elliptical shape of size 1$^{\circ}\times$1.8$^{\circ}$ around Sgr\,A$^{*}$.
Since then, several studies of the very central degrees ($\ell\pm 2^{\circ}$, $b\pm 1^{\circ}$) of the GC have been performed \citep[e.g.][]{koyama07,koyama09,yamauchi09,uchiyama11, nishiyama13,heard13sources,koyama18}. In many of those, the contribution of unresolved point sources to the 6.7\,keV line emission is accounted by scaling NIR data or the stellar mass distribution models (SMDs) based on NIR observations to the ridge X-ray emission, where unresolved point sources are producing almost all the extended hard X-ray emission \citep[][]{revnivtsev09}.
All studies agree, however, that after the subtraction of unresolved point sources with this method, in the central degrees at the GC, there remains a hard X-ray emission excess. This excess has been interpreted so far as a strong indication for the existence of very hot plasma in the GC, as a completely new population of sources \citep[e.g.][]{uchiyama11, nishiyama13,yasui15}, or as a fractionally larger population of already existing types of sources such as intermediate polars (IPs) \citep[e.g.][]{heard13sources}.

Regarding the physical explanation of excess Fe~XXV emission in the GC, various hypotheses have also been proposed. 
For example, \citet{belmont05} suggested that the very hot plasma could exist, if the medium is collisionless, as gravitationally confined helium plasma, since the hydrogen would have already escaped the GC region, while \citet{belmont06} propose a viscous heating mechanism in order to heat and maintain the plasma. \citet{uchiyama11} noticed that although the observed Fe~XXV excess is explained with difficulty by point sources, its flux distribution is similar to the shape of the NIR distribution of the nuclear stellar cluster (NSC) and nuclear stellar disc (NSD), yielding a connection to the point sources. They proposed that the plasma could be the result of multiple supernova explosions, which could be explained by the high density of molecular gas and the on-going star formation of the NSC and NSD.
\citet{uchiyama13} estimated the thermal energy and the dynamical age of the high-temperature plasma (5-8 keV) to be $1\times 10^{53}$\ergs{} and $2\times 10^4$\,yr respectively. They consider the supernova scenario highly unlikely since the required supernova explosion rate of $>5\times 10^{-3}$\,yr$^{-1}$ is too high to be explained by the stellar mass of the GC, and the fact that supernova remnants (SNRs) at ages of $\sim$10$^4$\,yr have significantly lower temperature than that of a plasma at 7\,keV. They rather propose an alternative scenario of many violent flares of Sgr A*, also suggested by \citet{koyama96}.
\citet{nishiyama13} used near-infrared polarimetric observations of the central $3^{\circ} \times 2^{\circ}$ of the Galaxy, and suggested that the diffuse thermal plasma is possibly confined by a large-scale toroidal magnetic field. With this explanation the required energy input is reduced by orders of magnitude. 
In addition, they propose other possible heating mechanisms of the gas, such as past activity of the super-massive black hole \citep{koyama96}, magnetic reconnection \citep[][]{tanuma99}, and star formation and consequent supernova explosions \citep{crocker12}. \citet{heard13sources}, analysing \xmm{} data of the GC, attribute the excess either to a different kind of underlying source population, or to an inaccurate SMD. They suggest that the Fe~XXV excess could be reproduced if the population of IPs is 7 times higher in the centre than other regions. More recently, \citet{oka19} analysed infrared spectra towards ~30 bright stars close to Sgr\,A$^{*}$ and found that the presence of  warm gas dominates the volume of the central molecular zone (CMZ). They claim that a very hot X-ray-emitting plasma could not coexist with the warm gas since it would have been cooled by the latter. They conclude that the very hot gas does not exist over extended regions and most probably the excess observed is due to unresolved stars and to the scattering of stellar X-rays by interstellar matter.

Overall, the existence of truly diffuse very hot plasma in the central degrees of our Galaxy is still debated, and even if it exists its origin is far from clear. A high-temperature component has also been revealed through the detection of the 6.7\,keV emission line in the core of M82, but its properties are difficult to constrain due to its low emissivity \citep[][]{strickland07,strickland09}. Investigating the existence and studying the physical properties of the 6.7\,keV emission in our own Galaxy will give valuable insights on what could be the driving mechanism of the outflows observed today in our and in other galaxies.

In this work we use all available \xmm{} observations of the GC and Galactic disc included within $\ell = \pm10^{\circ}$ and $b=\pm 2.0^{\circ}$. These observations provide the most detailed view of the Fe~XXV line emission so far, covering also the \chandra{} deep region \citep{revnivtsev09}. 
To estimate the contribution of point sources to the Fe~XXV line emission we use SMD models based on photometric as well as kinematic data and compare our results with an NIR \spitzer{} map.
In Sect.\,\ref{data} we describe the \xmm{}  and \spitzer{} data used in this work, and their corresponding analysis, as well as the SMD models.
In Sect.\,\ref{analysis} we present latitudinal and longitudinal profiles of the 6.7\,keV emission line, SMD and NIR data, and calculate the excess iron 6.7\,keV line emission in the GC.
In Sect.\,\ref{modelling} we model and calculate the physical properties of the excess Fe~XXV emission, whereas in Sect.\,\ref{discussion} we discuss our results in tandem with other works and suggest physical explanations for the observed excess. Finally, in Sect.\,\ref{summary} we summarise our findings.
Throughout this work we use a distance to the GC of $\mathrm{8.2\,kpc}$ \citep[][]{gravity19}, errors are reported at the 1$\sigma$ level unless otherwise stated, and acronyms are summarised in the appendix Table \ref{tab.acronyms}.

\section{Data}\label{data}

In the following section we describe all data (X-ray and infrared) as well as the SMD models used in this work. 

\subsection{The \xmm{} X-ray mosaic}\label{xraymosaic}

We have used a total of 370 \xmm{} observations with more than 6\,Ms clean exposure time (EPIC-pn equivalent), which comprise all available observations until April 1 2022, covering the central degrees of the GC and disc (out to $b=\pm2.0^\circ$, $\ell=\pm10^\circ$). 
Observations with less than 5\,ks have been masked out.
This sample includes all observations of the GC reported by \citet{ponti15} and \citet{ponti19}, as well as all newer GC observations. Moreover, it includes all serendipitous observations along the Galactic plane, as well as 46 observations from the Heritage programme (ID: 088393) that will eventually map the central 40 square degrees of the Galactic plane. In the appendix Table \ref{tab.xmm} we present all \xmm{} observations (237) not already presented in detail in \citet{ponti15}.

We reduced and analysed all EPIC observations, using the \xmm{} Science analysis system (SAS) v19.0.0. We followed the same procedure presented in detail by \citet{haberl12} and \citet{ponti15}. Briefly, we produced calibrated event files using the \textit{emchain} SAS task for the MOS cameras and \textit{epchain} SAS task for the pn camera. The latter was also used to create out-of-time event files for the pn camera using \textit{withoutoftime=Y}, in order to subtract out-of-time events and properly correct for the charge transfer inefficiency.
We also used the \textit{emtaglenoise} tool to flag noisy MOS CCDs at low energies \citep{kuntz08}.
In order to filter background flares we created good time interval files using the SAS task \textit{tabgtigen} with a constant cut-off of 2.5 cts\,s$^{-1}$ and 8.0 cts\,s$^{-1}$ for the EPIC MOS and EPIC pn exposures, respectively. In addition, we visually inspected all light curves and selected custom cut-offs when needed, as indicated in the appendix Table \ref{tab.xmm} for observations not reported in \citet{ponti15}. The final product was filtered event lists for each detector. 

We then created images and exposure maps for EPIC pn and MOS in the five standard bands of \xmm{} which are traditionally used to run source detection, using the clean event files and the tool \textit{eimageget}, as well as a band to represent the Fe~XXV emission line. Therefore, the bands used are: Band 1: $0.2-0.5$\,keV, Band 2: $0.5-1.0$\,keV, Band 3: $1.0-2.0$\,keV, Band 4: $2.0-4.5$\,keV, Band 5: $4.5-12.0$\,keV, and Fe~XXV Band: $6.62-6.8$\,keV. The EPIC pn Band 5 images were corrected for strong contaminating instrumental background emission due to Ni, Cu, and ZnCu lines, by removing the emission at energies 7.2--9.2 keV.
Then, the detector background created from filter wheel closed data was subtracted for each
observation \citep[for a detailed description see:][]{maggi16}.

In each band and for each detector, a mosaic was created combining all observations, after the subtraction of point sources detected in the \xmm{} observations, and strong stray-light artefacts. The method for the removal of stray-light artefacts is described in detail in Sect.\,2.2.1 of \citet{ponti15}.
The result was a background-subtracted map for each detector. We then combined the MOS and pn maps to a single count-rate map, where the exposure maps of the MOS detectors were multiplied by a scaling factor to account for effective area differences at 6.7\,keV. This procedure is described in detail in the appendix. 
Finally, the mosaic was adaptively smoothed  with a minimum signal-to-noise of 10, and following the standard procedure described in the \texttt{asmooth} tool documentation. We calculated the error of the smoothed mosaic using the command \texttt{readvariance=yes} of the \texttt{asmooth} tool after supplying a variance map. The supplied variance map was calculated after propagating the errors of the raw images and background maps following Gehrels approximation \citep[][]{gehrels86}.

\begin{figure*}
 	\centering
 		\includegraphics[scale=1.4]{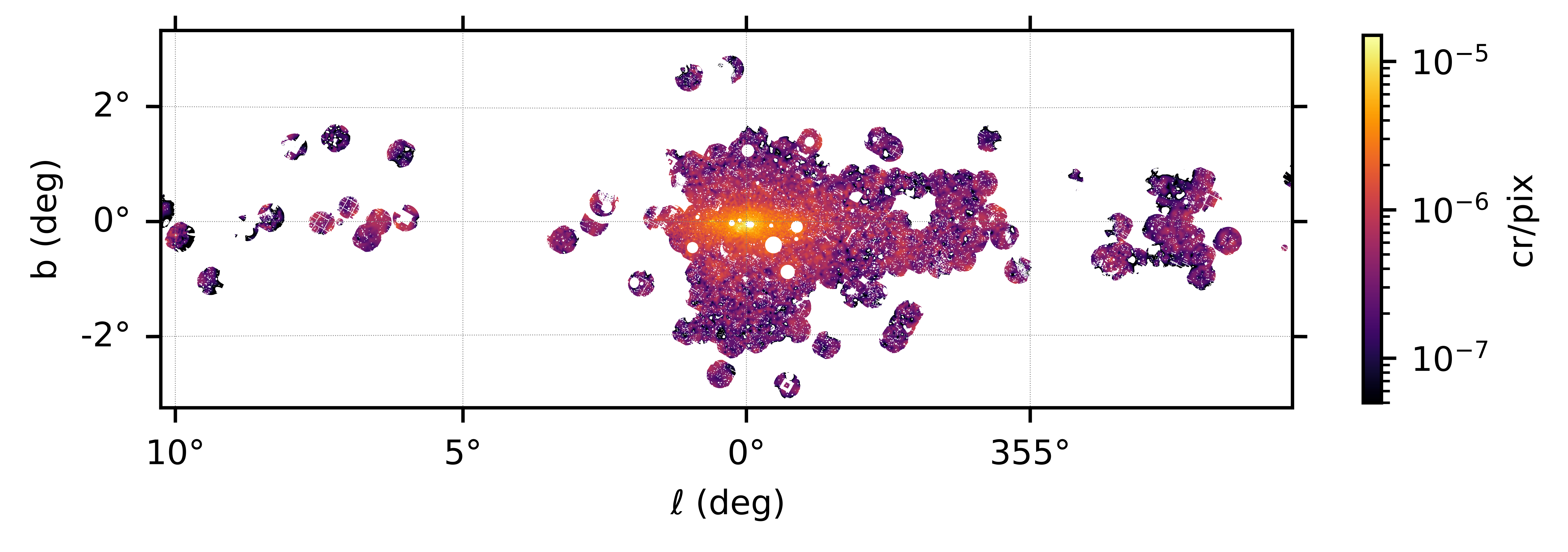}
 		\includegraphics[scale=1.35]{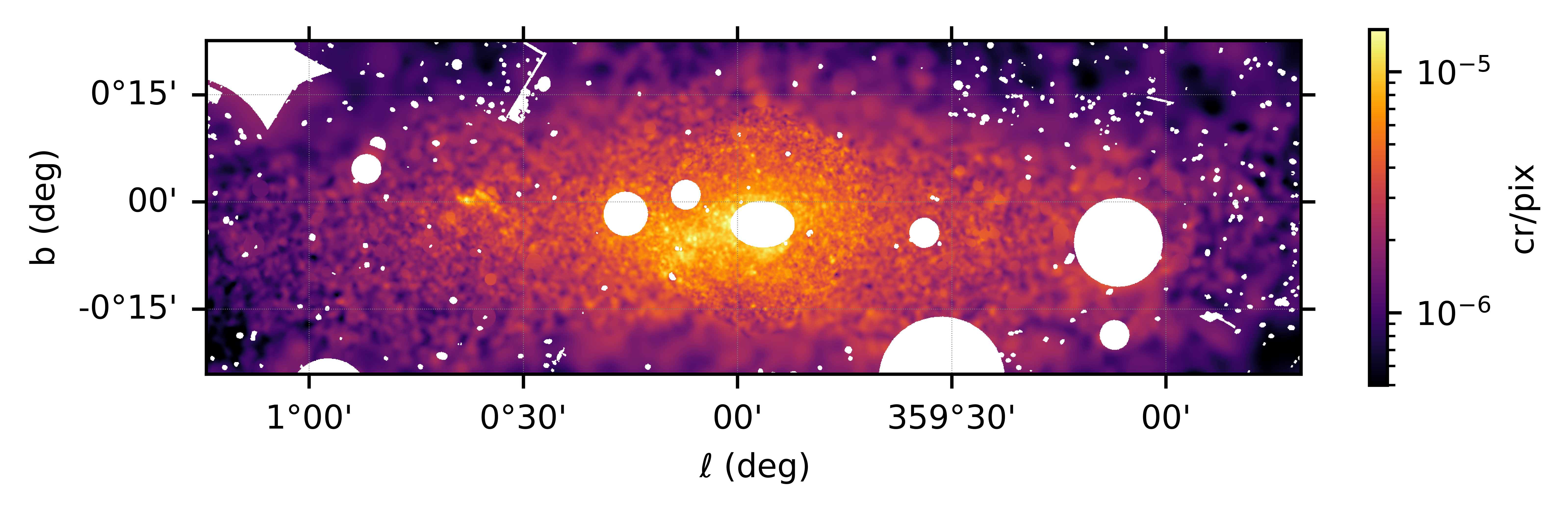}
 		\caption{\xmm{} 6.7\,keV emission from the GC and inner disc. Top: EPIC \xmm{} count rate mosaic in the Fe~XXV band. Bottom: The same as top but showing only the emission from the central degrees of our Galaxy. Regions containing bright sources are masked.}
 		\label{fig.xmm}
 
 \end{figure*}

Since in the very central arcmin of our Galaxy there are many bright sources (e.g. Sgr\,A East, Arches cluster, etc.), we removed larger regions in order to avoid contamination of the hard diffuse emission (6.7\,keV) by the scattered light halos corresponding to these sources \citep[for an example of dust scattering halos around a bright source see:][]{jin2017}. We present the excised regions in  Table \ref{tab.extrasources}, while the \xmm{} mosaic with sources removed in the Fe~XXV energy band is shown in Fig.\,\ref{fig.xmm}.

\begin{table}
	\caption{Bright sources removed from the XMM-Newton X-ray mosaic}
		\begin{tabular}{@{}lccc@{}}
			\hline  
			Name & $\ell$  & $b$  & radius \\
			 & deg & deg  & arcmin \\
			 \hline
			1E 1740.7--2942 & 359.111 & -0.094 & 6.1 \\
			SLX 1744--299/300& 359.273& -0.891 & 7.0\\
			2E 1742.9--2929 & 359.504 & -0.414 & 8.5 \\
		    KS 1741--293 & 359.563 & -0.072 & 2.0 \\
		    Sgr A East & 359.941 & -0.052 & 3.14$\times$4.41 
		    \\
		    Arches cluster & 0.120 & 0.016 & 2.0 \\
		    1E 1743.1--2843 & 0.260 & -0.028 & 3.0 \\
		    G0.9+0.1 & 0.866 & 0.077 & 3.0 \\
		    
		    IGR J17497--2821&0.954 & -0.461 & 5.7 \\
		    
			\hline			
			\end{tabular} 	
		\label{tab.extrasources}
		\end{table}

\subsection{Stellar mass distribution models}\label{SMDs}

The \xmm{} Fe~XXV map presented in Fig.\,\ref{fig.xmm} shows the diffuse X-ray emission from the 6.7\,keV line produced by unresolved point sources as well as possibly by very hot plasma. The contribution of the unresolved point sources is expected to correlate with the stellar mass distribution in the Galaxy. Therefore, we compare the 6.7\,keV emission with SMDs of the Milky Way.

The total stellar density of the Milky Way can be conveniently decomposed as the sum of several components:
\begin{equation}
\rho_{\rm TOT}(x,y,z) = \rho_{\rm NSC} + \rho_{\rm NSD} + \rho_{\rm BAR} + \rho_{\rm DISC} \,.
\end{equation}
In order of increasing Galactocentric radius $R$, these components are: (i) The NSC, which is a dense, massive ($M \simeq 2.5 \times 10^7\, \msun$) and slightly flattened assembly of stars centred on Sgr\,A$^{*}$ \citep{schoedel14,neumayer20}. It dominates the stellar mass density at $R\lesssim10\pc$. (ii) The NSD, which is a flattened stellar structure with a mass of $M\simeq 1.05 \times 10^9\, \msun$ \citep{launhardt02,sormani22nsd} that dominates the mass density at Galactocentric radii $10 \lesssim R \lesssim 200\pc$. (iii) The Galactic bar, which is a strongly non-axisymmetric structure whose major axis lies within the Galactic plane, with its nearer end at positive longitudes (for a review see for example \citealt{Bland-Hawthorn2016}). It has a mass of approximately $M\simeq 1.9 \times 10^{10} \, \msun$ and dominates the stellar mass density in the range $0.2\lesssim R \lesssim3\kpc$. (iv) The Galactic disc, which dominates the stellar mass density at $R\gtrsim 3\kpc$.
We consider two models in this paper, Model 1 and Model 2. They are constructed by combining recent models of the above individual components, as summarised in Table \ref{tab:massmodels}. The NSD is the component that is most important for our present purposes, because it dominates the stellar mass density in the GC region shown in the bottom panel of Fig.~\ref{fig.xmm} (recall that 1 degree at the distance of the GC corresponds to $\simeq$140 pc, which is roughly the radius of the NSD). Model 1 includes the NSD fiducial model from \cite{sormani20}. The shape of the density profile of this model was previously derived by \citet{Gallego-Cano2020} from a S{\'e}rsic profile fitting to the Spitzer/IRAC 4.5\,$\mu m$ maps. Model 2 includes the NSD model from \cite{sormani22nsd}, which was fitted purely to kinematic data without using any photometric information. The two NSD models are therefore fitted to different data sets using different methods. Therefore, they give two independent assessments of the stellar mass density profiles of the NSD.

\begin{table*}
\caption{Stellar mass distribution models of the Milky Way used in this paper}
\centering
\scalebox{0.93}{
\begin{tabular}{lll}  
\toprule
 &	Model 1 & Model 2 (fiducial) \\
\midrule
Nuclear Star Cluster & Best-fitting model from \citet{chatzopoulos15} & Best-fitting model from \citet{chatzopoulos15}  \\
Nuclear Stellar Disc & Model 3 of \citet{sormani20}  & Fiducial model of \citet{sormani22nsd} \\
Galactic bar & Best-fitting model of \citet{launhardt02} & Bar + long bar from \citet{sormani22bar} \\
Galactic Disc & Best-fitting model of thin+thick disc from \citet{mcmillan17} & Disc from \citet{sormani22bar} \\
\bottomrule
\end{tabular}}
\label{tab:massmodels}
\end{table*}

\subsection{Near-infrared maps}

For the estimation of the stellar component in the 6.7\,keV emission we have additionally used a NIR map. 
The NIR emission (3-4 $\mu$m) has been found to be a good tracer of the stellar density and to scale with the Fe~XXV line emission \citep[e.g.][]{revnivtsev06a,revnivtsev06b}.
For that reason we have constructed a NIR map using \spitzer{} data, which we use only as comparison to our SMD models. The proper usage of the \spitzer{} map would require various corrections (e.g. for foreground stars, supergiant contribution, etc.).

Then, we built for the central 5 square degrees of the GC, \spitzer{} \irac{} Band 1  (3.6$\mu$m) and Band 2 (4.5$\mu$m) mosaics, using the toolkit \textit{Montage}\footnote{http://montage.ipac.caltech.edu/}. 
For the creation of the mosaics more than 1000 archival observations were used each time, which then were re-projected with the tool \texttt{mProjExec} and finally corrected for background differences.

Since extinction towards the GC at this wavelength is non-negligible, we corrected the \spitzer{} maps using the extinction map provided by \citet{schoedel14} for the 4.5 $\mu$m band. This map is the best choice in terms of resolution (5 arcsec), since it is based also on \spitzer{} data. However, since it does not cover our entire \irac{} mosaic, we decided to show only the central region ($1^\circ\times 1^\circ$) with a reliable extinction measurement \footnote{There are other maps, which we decided not to use because of their lower resolution. For example the extinction map provided by \citet{schultheis14} in the $K_s$ band and up to a distance of 8\,kpc, is the best option in terms of coverage but has a resolution of 6 arcmin. Also, \citet{marshall06} provide a 3D map  of the entire Galaxy based on \mass{} data, but with lower resolution (15 arcmin), and \citet{green19} provide a 3D extinction map for the entire Galaxy, based on Gaia parallaxes and stellar photometry from \textit{Pan-STARRS} 1 and \mass{}. However, towards the GC, the extinction could not be calculated for distances greater than $\sim$3 kpc due to the lack of data.}.

We have created the mosaics in both bands (3.6 and 4.5\,$\mu$m) in order to assess differences that could be attributed to the differential extinction in these two bands. We noticed that after extinction correction, the differences between the two wavelengths were negligible. We decided to use the 4.5$\mu$m map for the rest of this work because it is generally less affected by extinction compared to the 3.6\,$\mu$m band.

\section{Analysis and results}\label{analysis}

In this section we report on the analysis performed on the \xmm{}, and \spitzer{} data, along with SMD models, in order to produce a map representing the Fe~XXV line emission in excess to what is produced by unresolved point sources. As a first step, since all maps have different orientation and resolution, we used the \texttt{astropy} module \texttt{reproject} of python, to reproject the corrected \spitzer{} map, as well as the SMD models, to the same pixel size and orientation as the \xmm{} mosaic.

\subsection{Latitudinal and longitudinal profiles}\label{profiles}

\begin{figure}
 	\centering
 		\includegraphics[width=1.0\columnwidth]{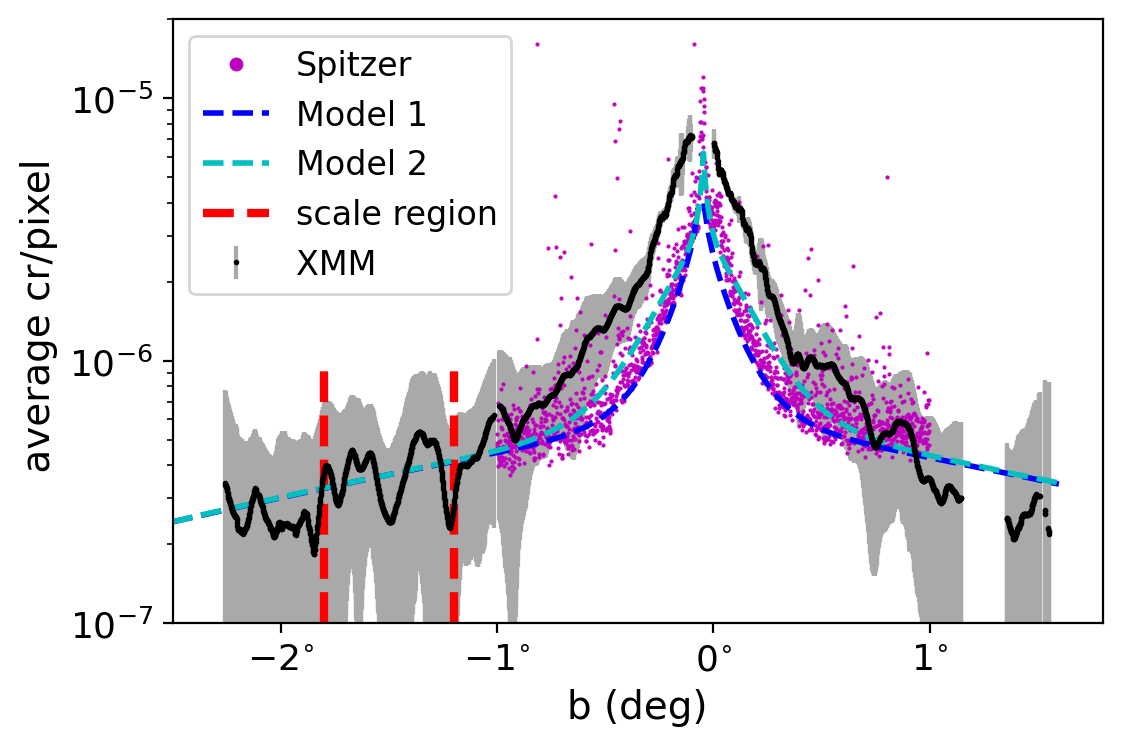}
 		\includegraphics[width=1.0\columnwidth]{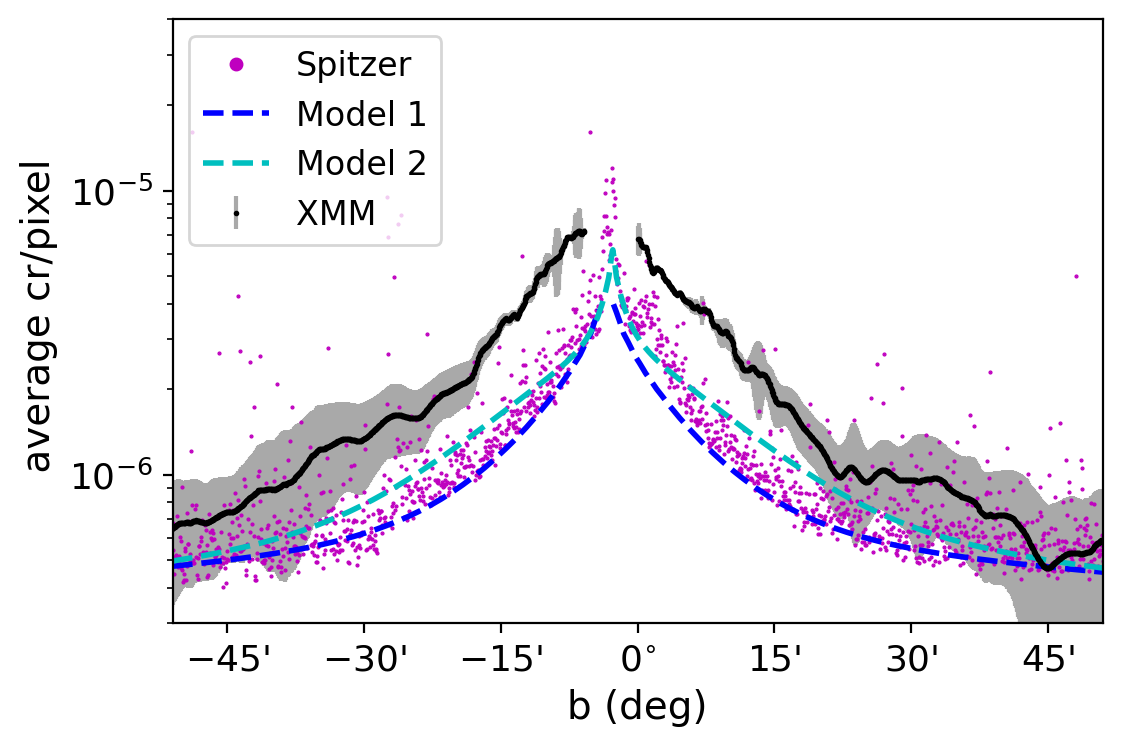}
 		\caption{Latitudinal profiles of the X-ray, SMD models, and infrared data. Top panel: Average count rate per pixel for the Fe~XXV Band over Galactic latitude ($b$), extracted from a profile of 0.5 deg width centred at Sgr\,A$^{*}$. The \xmm{}  and \spitzer{} data are shown with the black line and the magenta dots respectively. The profiles extracted from the SMD Model 1 and Model 2 are shown with blue and cyan dashed lines respectively. The region enclosed within the red dashed lines is used to scale the \xmm{} and SMD profiles. 
 		Bottom panel: Same as top but for $b$ between $\pm0.85^{\circ}$.}
 		\label{fig.profilelat}
 
 \end{figure}

In order to assess the existence of truly diffuse very hot plasma in the central degrees of the Galaxy, we extracted intensity profiles centred on Sgr\,A$^{*}$, with a width of 0.50$^{\circ}$. The profiles were extracted from the \xmm{} EPIC count rate image, the SMD models, and the \spitzer{} map,  spanning Galactic latitudes from  $b=-2.73^\circ$ to $b=+1.60^\circ$, and Galactic longitudes between $\ell=351.53^\circ$ and $\ell=+8.10^\circ$.
The SMD and \spitzer{} profiles were scaled to the average value of the area covered by the latitudinal profiles from $b=-1.2^\circ$ to $b=-1.8^\circ$ (indicated as scale region in top panel of Fig.\,\ref{fig.profilelat} and has an error of 5\%), under the assumption that all the diffuse emission originating from this region is due to unresolved point sources, since it includes the \chandra{} deep region \citep[][]{revnivtsev09}, and that the  X-ray emissivity over stellar mass density or NIR flux within the scale region remains the same over the entire profile (1:1 scaling).

In the top panel of Fig.\,\ref{fig.profilelat}, we show the X-ray and the scaled \spitzer{} and SMD profiles along Galactic latitude. At the very centre of our Galaxy we show no data for the X-ray profile since a large region around Sgr\,A East has been removed.
We note that the SMDs, as well as the \spitzer{} data, are in very good agreement with one another, even though no detailed corrections (regarding foreground stars, bright supergiants) have been applied in the latter case. 
There is a small difference ($\sim8\%$ on average) between Model 1 and Model 2, visible from $b\pm0.5^{\circ}$ to $b\pm0.2^{\circ}$ (see bottom panel of Fig.\,\ref{fig.profilelat}). This reflects systematic differences between the calculations of the two models, mainly the NSD component, since the first is based on photometric and the second on kinematic data. 
We see that no matter what method we use to account for the unresolved point sources there is always an excess of X-ray emission in the very central degrees. We thereafter use Model 2 as the fiducial model since it is based on more recent work, and has smaller errors than Model 1 (10\% versus 25\%).

\begin{figure}
 	\centering
 		\includegraphics[width=1.0\columnwidth]{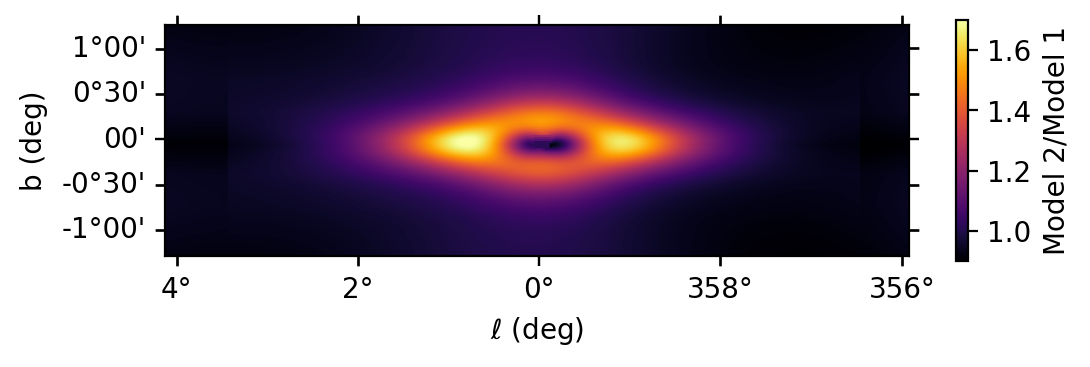}
 		\caption{Ratio of Model 2 over Model 1. Both models are scaled as explained in Sect.\,\ref{profiles}. }
 		\label{fig.ratiomodel}
 \end{figure}

When we calculated the longitudinal profile, Model 1 showed smaller values within $\rm{\ell<\pm2.0^{\circ}}$ (of the order of $\sim20-25\%$) than Model 2. This is probably due to the fact that the bar is thicker in Model 1 \citep[][]{launhardt02} than in Model 2 \citep[][]{sormani22bar}. Moreover, the bar model calculated by \citet{launhardt02} is based on photometric data from \textit{COBE DIRBE} that have a low angular resolution of $0.7^{\circ}$. Overall the bar/bulge models agree within the errors, since for Model 1 the average error is of the order of 25\%, while for Model 2 is of the order of 10\%. In Fig.\,\ref{fig.ratiomodel} we show the ratio of Model 2 over Model 1 within $\ell\pm4.0^{\circ}$ and $b\pm1.0^{\circ}$, where the difference between the bar/bulge model between the two models is apparent.

\begin{figure}
 	\centering
 		\includegraphics[width=1.0\columnwidth]{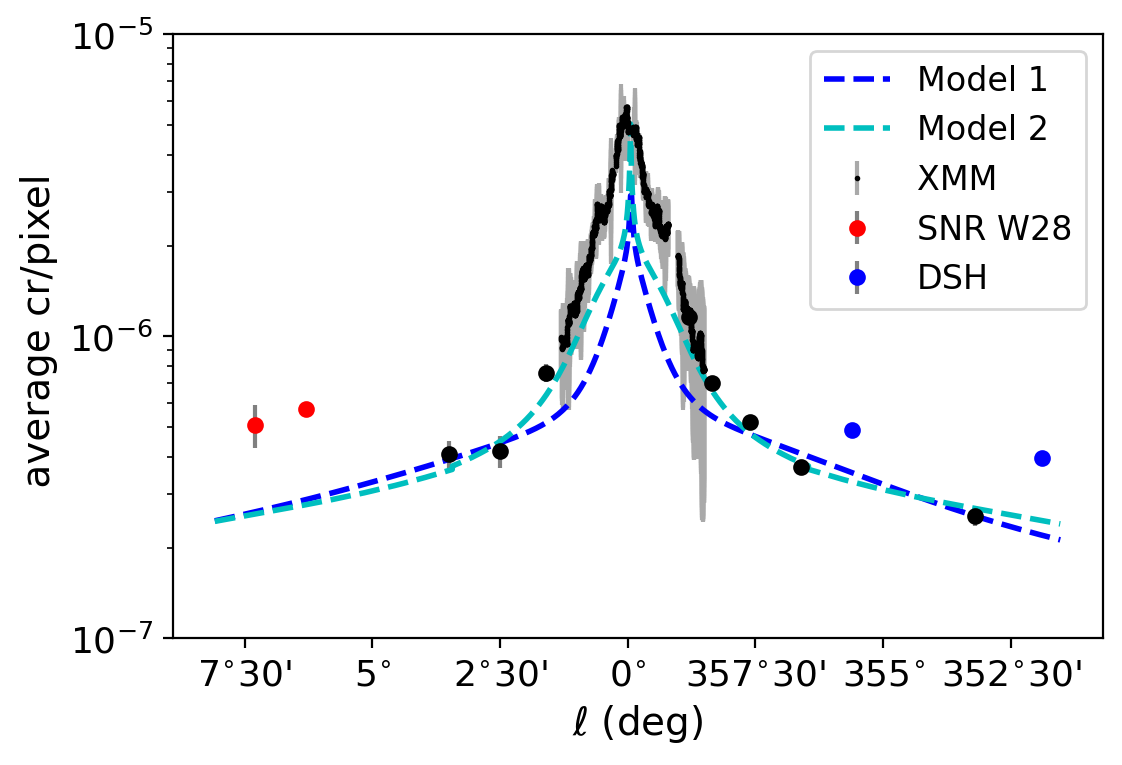}
 		\caption{Longitudinal profile of the X-ray, and SMD models data. Average count rate per pixel for the Fe~XXV Band over Galactic longitude ($\mathrm{\ell}$), extracted from a profile of 0.5 deg width, centred at Sgr\,A$^{*}$. The \xmm{} data are shown with black dots while the profiles extracted from the SMD Models 1 and 2 are shown with the blue and cyan dashed lines respectively. The red and blue points are \xmm{} observations that include the SNR W28 and the dust scattering halos from bright XRBs, respectively, and therefore show an excess compared to the scaled Model 2.
 	}
 		\label{fig.profilelong}
  \end{figure}

In Fig.\,\ref{fig.profilelong} we show the longitudinal profiles for the X-ray data, the fiducial 1:1 scaled Model 2, and Model 1.
For longitudes outside the GC ($\ell>4.0^\circ$\,and $\ell<-1.5^\circ$)
the available observations are either not continuous, and/or of lower total exposure time than the ones in the GC (see Table \ref{tab.xmm}). The non-continuous observations were typically made for the purpose of observing specific interesting sources such as SNRs, or X-ray binaries (XRBs), and not for the purpose of mapping the diffuse emission along the Galactic plane. For those observations, we decided to calculate the average intensity in one-degree sections within the profile wherever observations were available, while for the Heritage programme observations ($-1.0^\circ>\ell>-4.0^\circ$), we also averaged in steps of $1^\circ$. We show the average values of these larger regions with black circles in Fig.\,\ref{fig.profilelong}.
We see that the X-ray emission expected to originate from stars (1:1 scaled Model 2), is in good agreement with the \xmm{} data except for the very central degrees ($\ell$ within $\pm1.2^\circ$) and the red and blue points.
The observations at $\ell=6.0^\circ-7.5^\circ$ (red circles), are all covering the supernova remnant W28 \citep[e.g.][]{zhou14,okon18}, which are also visibly brighter in the \xmm{} mosaic (Fig.\,\ref{fig.xmm}).
This supernova remnant has an extent of about 1.5$^{\circ}$ and an absorbed flux of $F_{\rm{2-10\,keV}}=1.1\times10^{-12}$ \funit\ (see \chandra{} catalogue of SNRs\footnote{https://hea-www.harvard.edu/ChandraSNR/snrcat\_gal.html}). This results in a factor of 2.2 excess of the measured Fe~XXV emission compared to that expected from the stellar populations. The observations shown with  blue points correspond to those of the candidate supergiant fast X-ray transient IGR J17354--3255 ($\ell=-4.5^\circ$) and the accreting pulsar IGR J17255--3617 ($\ell=-8.5^\circ$).
Both of these sources, although removed from our final mosaic, show a dust-scattering halo (DSH) which contributes to a factor of 1.7 excess of the X-ray profile.
The black dot at around $\ell=-7.0^\circ$ corresponds to two observations, one of which is on the supernova remnant G352.7--00.1.
The reason we do not see an excess here, even though an SNR is observed, is probably because the SNR is quite small in extent ($\sim$10 arcmin), and has a much lower flux of $\mathrm{F_{2-10\,keV}}=6.6\times10^{-13}$ \funit, compared, for example, to W28.

\subsection{Calculation of the excess Fe~XXV emission}\label{calculationofexcessemission}

Since the X-ray emission shows an excess compared to the scaled 1:1 SMD model, we aim in this section to provide a quantitative view of this excess. For this reason we created a map of the central degrees of our Galaxy, covering the area where an excess is visible according to the profiles we presented in the previous section. The excess map was created by subtracting the scaled (in the same way as done for the profiles) 1:1 fiducial SMD map from the \xmm{} count-rate map. 

\begin{figure}
 	\centering
 		
 		\includegraphics[width=1.05\columnwidth]{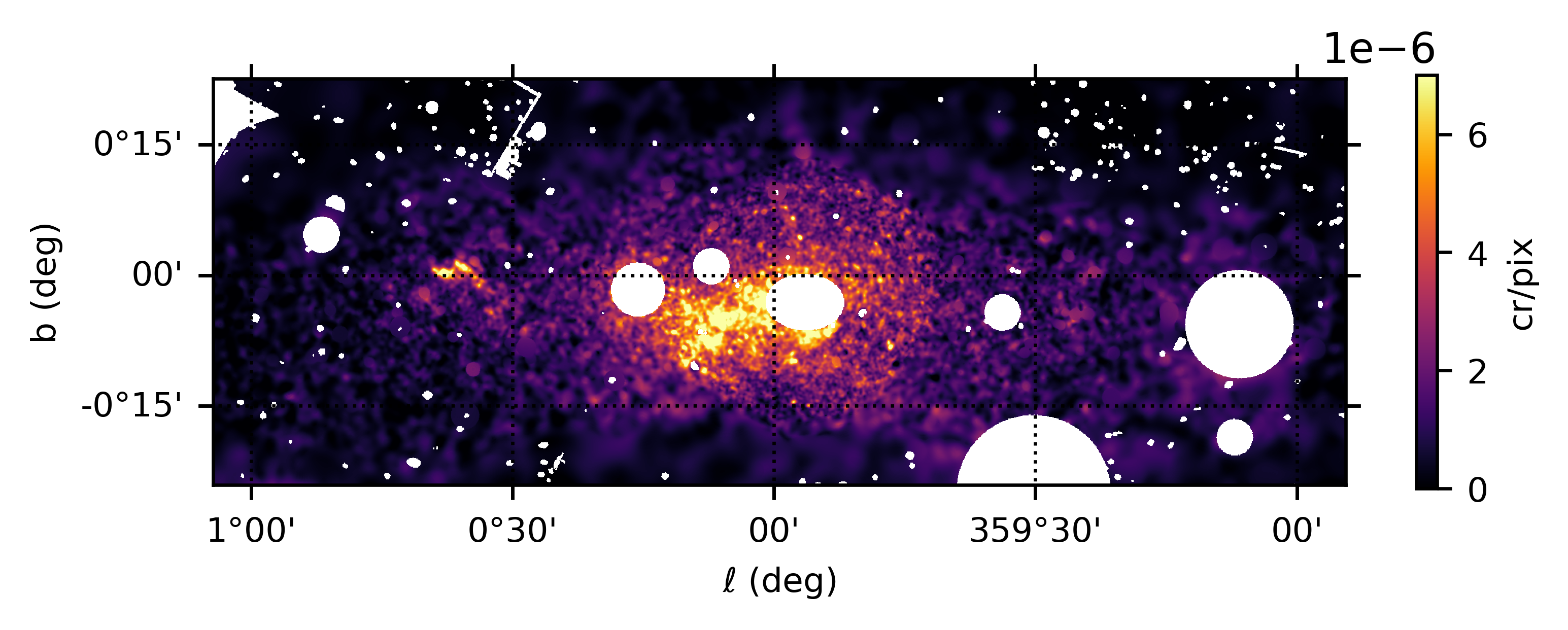}
 		\includegraphics[width=0.90\columnwidth]{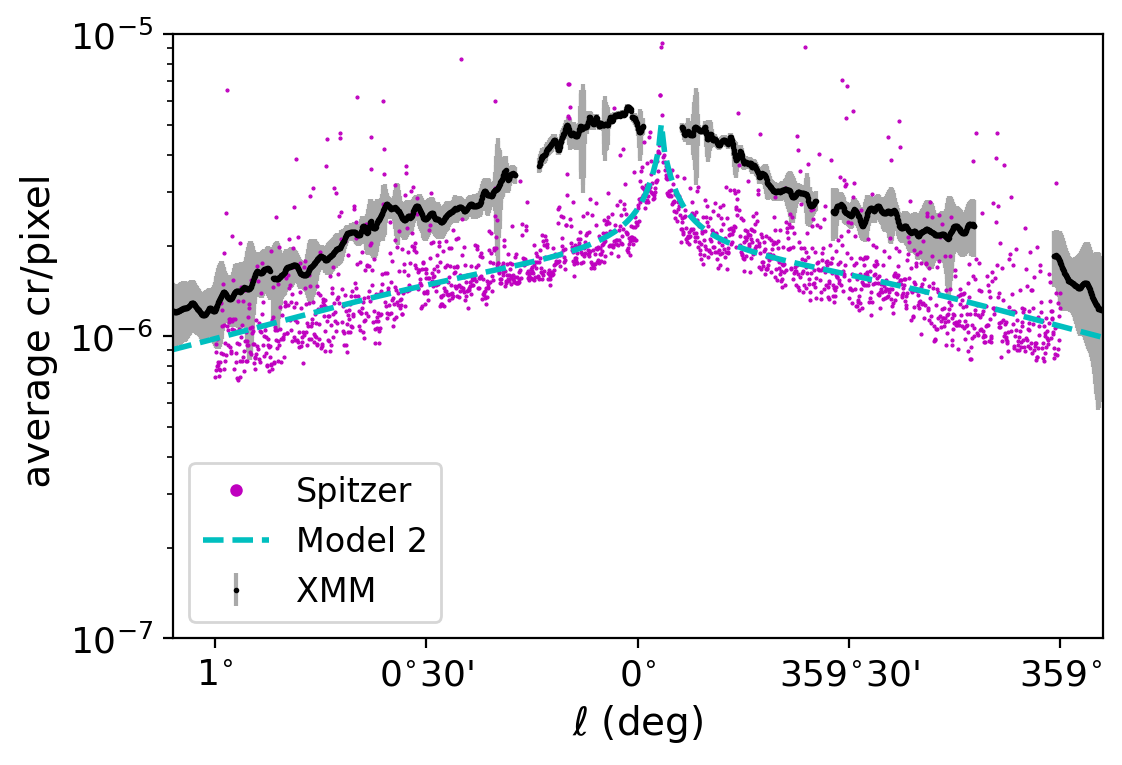}
 		\caption{Excess Fe~XXV emission in the GC and the corresponding X-ray, infrared, and Model 2 latitudinal profiles. Top: Fe~XXV excess emission attributed to diffuse emission after removing the unresolved point-source emission using Model 2 and 1:1 scaling.
 		Bottom :  \xmm{} longitudinal profile of the overall 6.7 keV emission covering the central 2$^{\circ}$ along with the scaled \spitzer{} and Model 2 profiles. Bumps on the longitudinal X-ray profile coincide with high-excess regions which are of known origin. For a detailed description see Sect.\,\ref{calculationofexcessemission}.
 		}
 		\label{fig.excessmaps}
  \end{figure}

We show in Fig.\,\ref{fig.excessmaps} the diffuse X-ray emission at 6.7\,keV which is in excess of what is produced by a 1:1 scaling of the unresolved point sources derived from the SMD Model 2.
The regions showing the higher excess correspond also to bumps on the longitudinal profile (Fig.\,\ref{fig.excessmaps}; bottom panel) that are recognised as known sources of X-ray emission. In particular, starting east of Sgr\,A$^{*}$ and going west we find the SNR G0.61+0.01 rich in 6.7\,keV emission \citep[e.g.][]{koyama07b,ponti15}.
Then, closer to Sgr\,A$^{*}$ an enhancement of diffuse emission has been observed with a number of distinct sources of X-ray emission, such as the pulsar wind nebula G0.13--0.11, the Quintuplet cluster ($\ell$=0.1604, $b$=-0.0591), and the candidate superbubble G0.1--0.1 as the dominant feature which shows a prominent Fe~XXV line in its spectrum \citep[][see their figures 2, 12 and 13]{ponti15}. At the same position as the candidate superbubble G0.1--0.1, the Sgr\,A molecular complex is located, which is very bright in the 6.4\,keV emission due to reflection.

Finally, an enhancement of diffuse emission is visible in our observations to the west of Sgr\,A$^{*}$ around the excised high-mass black-hole XRB 1E 1740.7--2942 \citep[e.g.][]{stecchini20}. 
This is likely to be either residual contaminating emission from the XRB or emission from the SNR  G359.12--0.05, which has a extent of $\mathrm{24\times 16}$ arcmin \citep[][see their figure 6]{ponti15}. \citet{nakashima10} studied the \suzaku{} spectrum of SNR G359.12--0.05 and found no clear sign of the Fe~XXV emission line, but they do see a hint of residual excess in the hard band.

Apart from the high-excess regions, which point to the existence of truly diffuse hot plasma similar to the case of the SNR G0.61+0.01, the emission seems to be more uniformly distributed, which could be the result of an older SNR population that has diffused and merged with the ISM or unresolved sources that for some reason are not accounted for by the SMD. Overall, the diffuse emission forms an ellipsoidal shape spanning two degrees in longitude and half a degree in latitude (see top panel; Fig.\,\ref{fig.excessmaps}), in agreement with what has been found in previous works \citep[e.g.][]{yamauchi93}.

\subsubsection{Contamination from reflection emission in the GC}\label{reflection}

In the CMZ region, X-ray reflection emission originating from dense molecular gas due to past activity from the supermassive black hole Sgr\,A$^{*}$ is bright in the 6.4\,keV band \citep[e.g.][]{ponti13,khabibullin22}.
In this section we examined whether the contribution of the reflection at the 6.4\,keV emission line could significantly affect our results in the 6.7\,keV band.

\begin{figure}
 	\centering
 		\includegraphics[width=1.05\columnwidth]{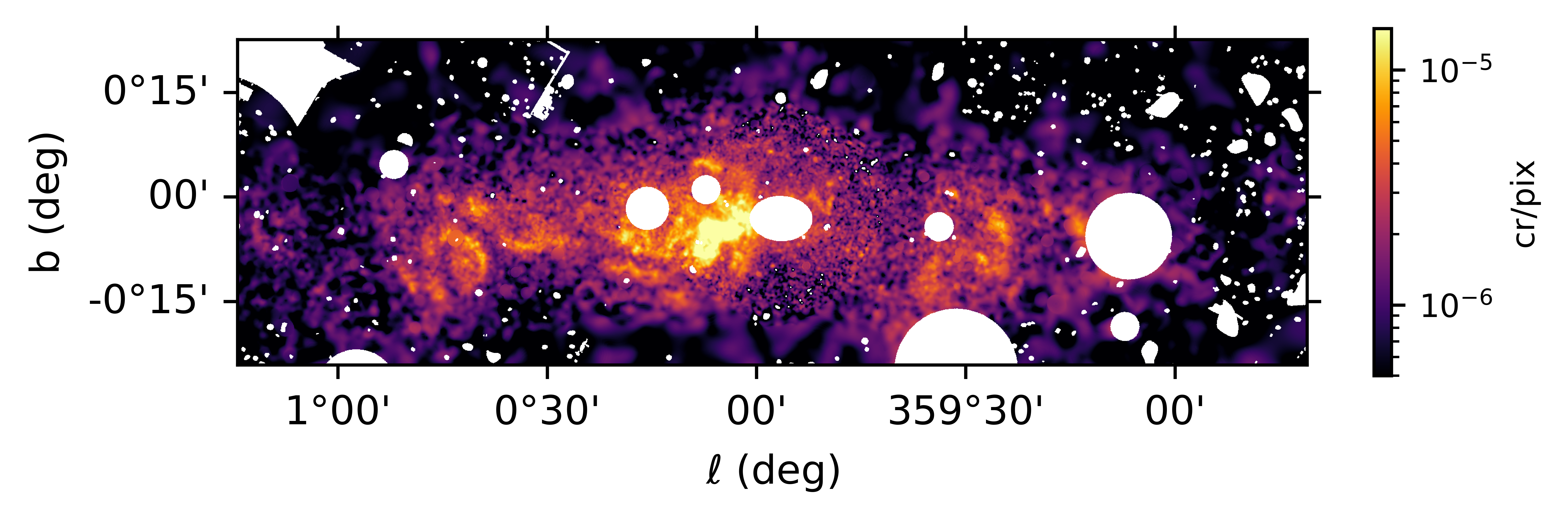}
 		\includegraphics[width=1.05\columnwidth]{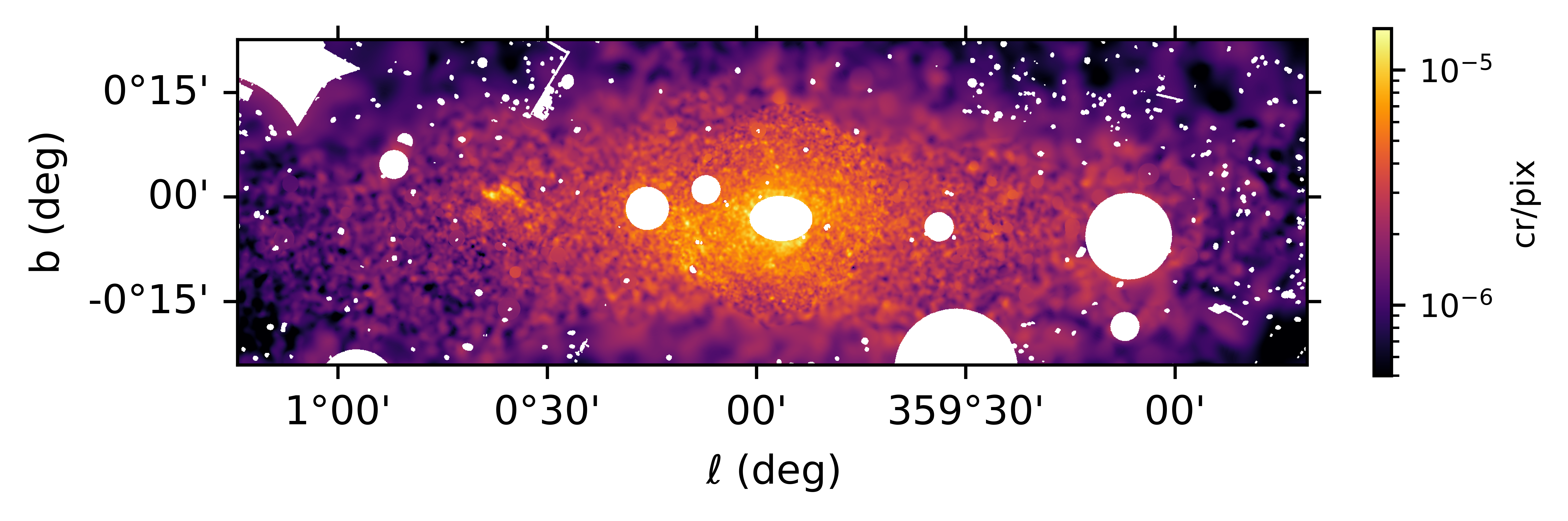}
 		\caption{Reflection and thermal emission in the GC. Top: Reflection emission (6.3-6.5\,keV band) in the central degrees of the GC. Bottom: Fe~XXV emission (6.62-6.8\,keV band) after correction for contaminating reflection emission.
 		}
 		\label{fig.reflection}
  \end{figure}

We started by creating an X-ray mosaic in the 6.3-6.5\,keV band to be representative of the reflection following the same analysis performed for the 6.7\,keV emission line map (see Sect.\,\ref{xraymosaic}).
Then we used a reflection model \citep[using an optical depth of the cloud of $\tau=0.5$ and a viewing angle of $\theta= 90^{\circ}$; for details see ][]{churazov17} and a thermal model (APEC; kT=7\,keV) and we folded them with an Auxiliary Response File (ARF) and an Redistribution Matrix File (RMF) of an observation close to the GC.
Under the assumption that the ARF and RMF do not change significantly along the X-ray map, we produced a simulated spectrum, quantified the contamination from reflection across the Fe~XXV map, and applied this correction to the 6.7\,keV map.
The top panel of Fig.\,\ref{fig.reflection} shows the reflection emission in the central degrees of the GC, and the bottom panel shows the Fe~XXV emission corrected for the contaminating reflection emission. We should note that the reflection emission is in fact variable, and in this section we present the time-averaged X-ray reflection signal since we are using all X-ray data available.

By comparing the uncorrected 6.7\,keV emission line mosaic with the one corrected for reflection (see bottom panels of Fig.\,\ref{fig.xmm}, and Fig.\,\ref{fig.reflection} respectively), we find that the contamination is not significant ($<$7\%) over the entire GC region, leaving the shape of the morphology of the 6.7\,keV emission practically unchanged. Although, for specific regions such as the Sgr\,A molecular complex ($\ell=0.110^{\circ}$, $b=-0.096^\circ$), the contamination can be up to 30\%, which explains part of the enhanced excess emission at that location in the top panel of Fig.\,\ref{fig.excessmaps}.

\subsubsection{Metallicity implications on the scaling of the SMD}\label{metallicityimplications}

In the previous sections, we calculated the excess Fe~XXV under the implicit assumption that the Fe~XXV emission and the stellar density scale in a linear way (1:1) along the entire Galaxy. 
However, different physical properties (i.e. metallicities) of the stellar populations of the NSD or NSC, for example, could result in a non-linear scaling between the SMD and the X-ray emission, and could possibly explain the enhanced Fe~XXV emission we observe in the GC. 

Indeed, \citet{schulteis21} using a KMOS (VLT, ESO) spectroscopic survey of K/M giant stars \citep{fritz21}, found that the global iron abundances ([Fe/H]) of the  NSC and the NSD are higher than that of the bulge (see their figure 8). In particular, their table 1 implies that the mean iron abundances of the NSD and NSC compared to that of the bulge are 1.35 and 1.65 times higher respectively. Moreover, \citet{krause22} studied the stellar populations in the transition region of the NSC and NSD, and found higher metallicities than those presented in \citet{schulteis21} and \citet{fritz21}, although their results are consistent within the errors.

To understand the effect that the higher metallicity in the NSC and NSD would have on the expected Fe~XXV emission we assumed that the Fe~XXV emission line can be represented well by a thermal plasma with a temperature of $\sim$7\,keV \citep[e.g.][]{koyama18}. For that reason, we varied the iron abundance of a \texttt{vapec} model in XSPEC according to the \citet{schulteis21} average values for the bulge, NSD, and NSC. We find that the differences in metallicity translate into roughly the same Fe~XXV emission enhancement for the NSD and NSC, namely $\times 1.25$ and $\times 1.52$, compared to that of the bulge/bar. In our calculations using XSPEC we assumed the abundance table provided by \citep{wilms00}. We have tested how our results change when using other abundance tables available in XSPEC and we found that they never vary by more than 2\%.

To quantify this, we calculated the 6.7\,keV flux excess when using the scaled 1:1 SMD and the non-linear ([Fe]) scaled SMD using higher metallicities for the NSD and NSD model components. In Table \ref{tab.flux} we present flux measurements for an elliptical region centred on Sgr\,A$^{*}$ with a minor axis of 1 degree along Galactic latitude and major axis of 3 degrees along Galactic longitude. We find that in this region reflection is contributing 9\% to the 6.7\,keV flux, while the resolved SNR G0.61+0.01 less than 0.5\%. 
Unresolved point sources contribute 65\% to the 6.7\,keV flux if we assume a 1:1 scaling of the SMD, while when we use the [Fe] scaling the unresolved point sources account for 75\%.
Therefore, for both scalings (1:1 and [Fe]) of the SMD, there is a 35\% and 25\% of the 6.7\,keV emission which still cannot be explained by unresolved point sources. In other words the hard X-ray emission we observe in the central degrees of the GC is 1.5 and 1.3 times greater than expected using the 1:1 and [Fe] scaling of the SMD, respectively.
We calculate that in order for all the Fe~XXV emission to be explained solely by metallicity enhancement of the unresolved point sources, a $\sim$1.9 times higher iron abundance for the NSC and NSD compared to that of the bulge/bar would be required. 
In Table \ref{tab.flux} we also show the total flux for Model 1 (1:1 scaling) just for comparison purposes with previous works \citep[e.g.][]{uchiyama11,heard13sources} that mainly have used SMD models based on photometric data such as the models of \citet{launhardt02}. In that case the 6.7\,keV emission in the GC is 1.9 times greater than what is expected from the 1:1 scaled Model 1 (while for Model 2 it is 1.5 times).

\begin{table}
	\centering
	\caption{Flux measurements within $b=\pm0.5^{\circ}$ and $\ell=\pm1.5^{\circ}$ }
		\begin{tabular}{@{}cc@{}}
			\hline  
		Map/Location	& $\rm{F_{6.7\,keV}}$   \\
		& $10^{-11}$\funit{} \\
		\hline
		\xmm{} & 4.81$\pm$0.01\\
		\xmm{} (corrected for reflection)& 4.41$\pm$0.02\\
		Model 1 (1:1 scaling) & 2.30$\pm0.46$ \\
		Model 2 (1:1 scaling) & 2.88$\pm0.30$ \\
		Model 2 ([Fe] scaling) & 3.30$\pm0.35$\\
		resolved SNR &  0.015$\pm$0.001 \\
		\hline
		\xmm{} excess (1:1 scaling) &1.52$\pm0.30$\\
		\xmm{} excess ([Fe] scaling) & 1.10$\pm0.35$ \\
			\hline		
			\end{tabular} 	
		\label{tab.flux}
		\end{table}

\section{Modelling of the excess Fe~XXV emission}\label{modelling}

In the following section, we model and estimate the physical properties of the excess Fe~XXV emission based on both the 1:1 and [Fe] scaling Model 2. We also make a uniform 7\% correction for the contribution of reflection, while we do not subtract the contribution from the known SNRs. We present models of the intensity and density distributions that attempt to explain the morphology we observe.
For all cases we fitted the images using the python \texttt{scipy} \texttt{curve\_fit} module. For the fitting of the images to the various models, a binning of 3 pixels (angular size of 12 arcseconds) was preferred, in order to increase the S/N and have more reliable statistics. 

\subsection{Model of the intensity distribution}\label{2d}

We modelled the two-dimensional (2D) intensity distribution of the excess emission using a power-law model which is described as 

\begin{equation}
I (x,y)=I_0\left[1+\left(\frac{x-x_c}{x_s}\right)^2+\left(\frac{y-y_c}{y_s}\right)^2\right]^{-a}
\label{eq.flatpo}
\end{equation}
where $I_0$ is the peak of the excess intensity which is located at $x_c$, $y_c$, x and y are the 2D coordinates, $x_s$ and $y_s$ are the scale heights along the x and y 2D coordinates respectively, while $a$ is the slope of the excess X-ray  intensity distribution.

Then, we also used a Sérsic model \citep{graham01} to investigate possible similarities between the excess X-ray emission and the NIR emission from the stellar populations of the NSD \citep{Gallego-Cano2020}. Therefore, we used exactly the same model as reported in \citet{Gallego-Cano2020}, namely:

\begin{equation}
I(x,y)=I_e \exp\left(-b_n\left[\left(\frac{p}{R_e}\right)^{1/n}-1\right]\right).
\label{eq.sersic}
\end{equation}

We note that $I_e$ is the excess X-ray intensity at radius $R_e$ where 50\% of the light is enclosed, x and y are the 2D coordinates and $p$ is defined as $p=x^2+(y/q)^2$, with q being the ratio between the minor and major axes. $b_n=1.9992n-0.32$ following \citet{capaccioli87} for $1 < n < 10$, as mentioned in \citet{Gallego-Cano2020}.
The best-fit parameters are presented in Table \ref{tab.2d}. The location of the peak X-ray excess is found to be at $x_c=0.011\pm0.015^\circ$, $y_c=-0.065\pm0.004^\circ$ while for comparison the location of Sgr\,A$^{*}$  is  at $\ell=359.94^\circ$, $b=-0.046^\circ$.

\begin{table*}
	\centering
	\caption{2D intensity models}
	\scalebox{0.93}{
	\begin{tabular}{@{}ccccc@{}}
			\hline  
		Components	&power-law [1:1] & power-law [Fe]  & Sérsic   & Sérsic [Fe]  \\ \\
		\hline
		$I_0$;$I_e$ ($\times 10^{-5}$cts/s/pixel) &5.95$\pm$ 0.01 & 5.33$\pm$ 0.01  &    1.08$\pm$ 0.01  &    1.06$\pm$ 0.01 \\

		$x_s$ (arcmin)	& 10.47$\pm0.04$&10.85$\pm0.07$ &- &- \\
		$y_s$ (arcmin)	&5.19$\pm0.02$&5.34$\pm0.02$&-&- \\
		$\alpha$  &0.718$\pm0.001$&0.810$\pm0.004$&- &-\\
		q	&-&-&$0.478\pm0.001$ &$0.476\pm0.001$\\
		n	&-& -&1.09 $\pm$ 0.01&1.03 $\pm$ 0.01\\
		$R_e$ (degrees)&	-&-&$0.52\pm0.01$&$0.46\pm0.01$\\
			\hline
			\chisqr (\chisq/dof) &1.10 (179398.41/162105) & 1.08 (176376.064/162105) & 1.13 (183317.45/162105) &1.10 (178908.49/162105) \\
			\hline		
			\end{tabular}} 	
		\label{tab.2d}
		\end{table*}

\begin{figure}
 	\centering
 		\includegraphics[width=1.0\columnwidth]{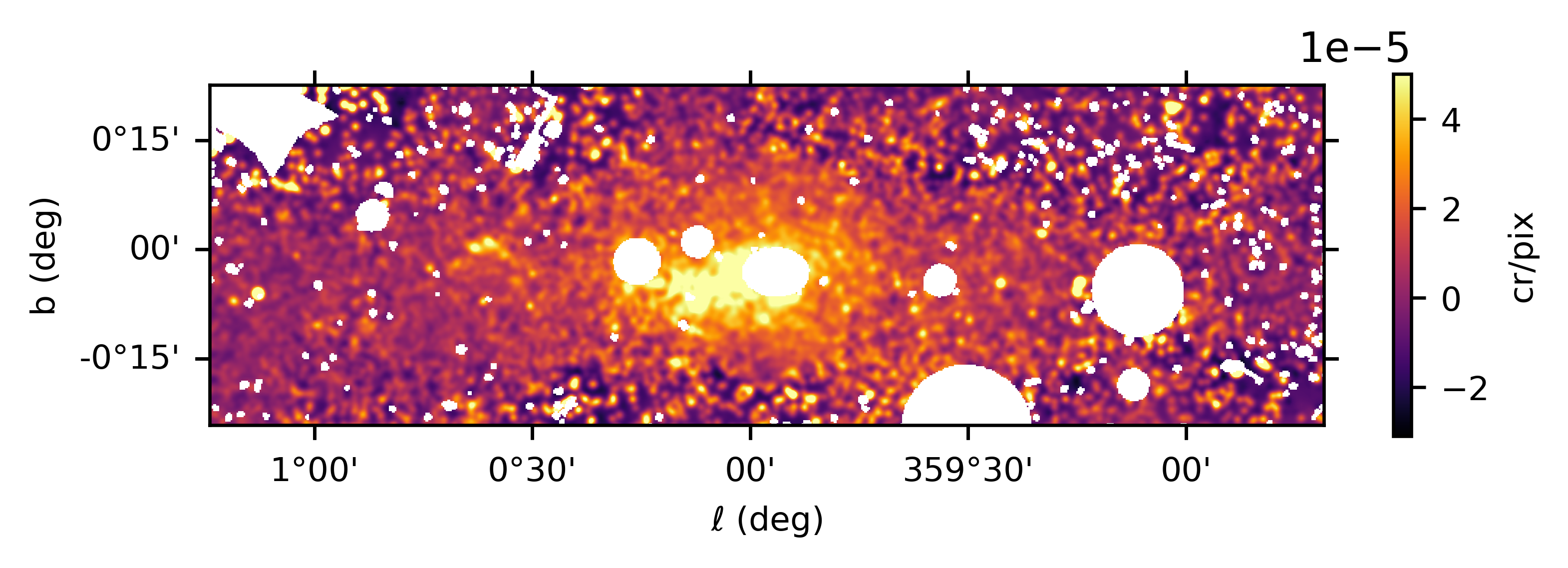}	
 		\includegraphics[width=1.0\columnwidth]{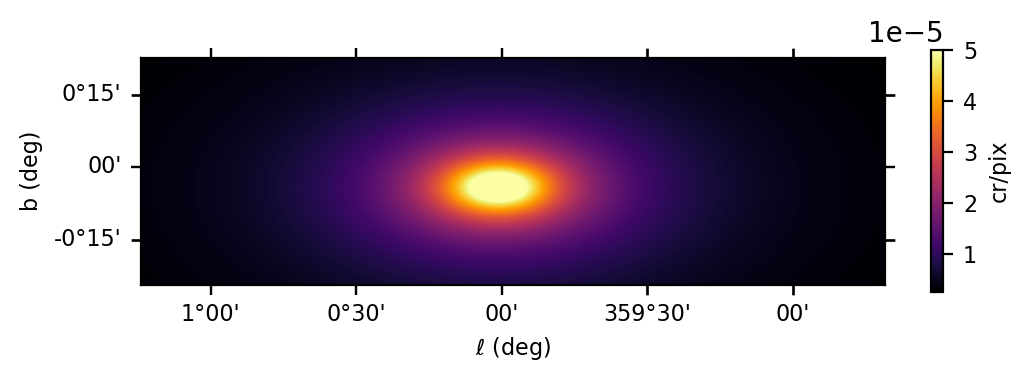}
 	    \includegraphics[width=1.0\columnwidth]{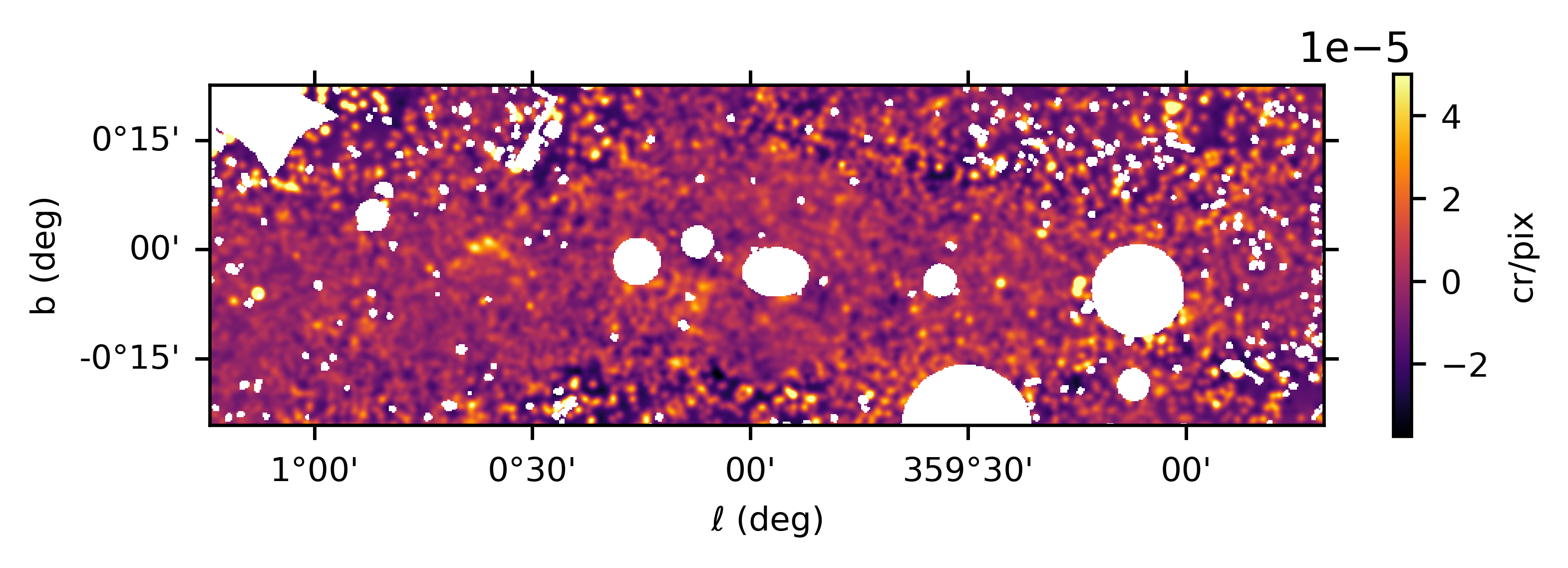}
 		\caption{Modelling of the X-ray intensity distribution in the GC. Top: Same as top panel of Fig.\,\ref{fig.excessmaps}, but with a larger pixel size ($12\times12$ arcsec) to allow for better statistics. Middle: Best-fit power law model for the distribution of the excess Fe~XXV emission. Bottom: Data minus model residuals.}
 		\label{fig.model_intensity}
 \end{figure}

In Fig.\,\ref{fig.model_intensity} we show the central $2^{\circ}\times 0.5^{\circ}$ of the \xmm{} count-rate excess emission mosaic (top panel), the best-fit model (middle panel) using a power-law model, and the data minus model residuals (bottom panel). The Sérsic best-fit model and residuals images are almost identical, therefore we do not show them. 
Both models provide an equally good fit (see Table \ref{tab.2d}) for the distribution of the excess X-ray emission.

Comparing the best-fitting parameters of the Sérsic model for the excess X-ray emission with those obtained from \citet{Gallego-Cano2020} for the NSD NIR emission, we find that our value for $n$ is smaller (1.03-1.09 versus 2.0-2.59), and the ratio of minor over major axis, $q$, in our case is also smaller (0.48 versus 0.60-0.85). Moreover, \citet{Gallego-Cano2020} find that 50\% of the NSD emission is within a radius of $R_e=4.57-5.66\pc$. We measure that 50\% of the excess emission is located within $\sim0.50^\circ$ which, for a distance to the GC of 8.2 kpc, corresponds to $\sim$70\,pc. Therefore, the excess emission we measure, although it appears similar (ellipsoidal shape) with the emission distribution of the NSD from \citet{Gallego-Cano2020} (see their figure 7), is definitely much flatter and broader, extending to larger Galactic longitudes.

Fitting the excess X-ray intensity distribution provides valuable insights on the extent and distribution of the excess Fe~XXV emission as well as on the location of the peak emission. From the residuals of the model (Fig.\,\ref{fig.model_intensity}; bottom panel) we notice that the model fits well the excess emission, except from regions corresponding to the bumps on the X-ray profile (discussed also in Sect.\,\ref{calculationofexcessemission}).

\subsection{Model of the density distribution}

We then assumed that all of the excess Fe~XXV emission originates from truly diffuse hot plasma with constant temperature and with a density that decreases as a function of distance from the location of the peak emission.
Therefore, we created a three-dimensional (3D) grid along the 2D coordinates $x$ and $y$, and the line of sight $z$. Each element of this grid contains a  different value for the density, which can be described by:

\begin{equation}
n(x,y,z)=n_0\left[1+\left(\frac{x-x_c}{x_s}\right)^2+\left(\frac{y-y_c}{y_s}\right)^2+\left(\frac{z}{x_s}\right)^2\right]^{-\beta}
\label{eq.density_3d}
\end{equation}
where $n_0$ is the peak density at the centre of the emission ($x_c$, $y_c$),  $x_s$ and $y_s$ are the scale heights along the x and y coordinates respectively, while $\beta$ is the slope of the distribution.
Here the line-of-sight axis $z$ is defined to have its centre at the GC, and the same scale height as the $x$ axis.

In the following analysis we assume that the temperature of the plasma is constant and has a value of $\sim$7\,keV which results from fitting global \suzaku{} spectra of the GC region \citep{uchiyama13,koyama18}. We should note here however, that the $\sim$7\,keV value found in previous works does not account for the unresolved point source contribution, which could shift the plasma temperature. The required temperature to produce a 6.7\,keV line is of the order of 5-10\,keV. 

The intensity-density relation can be given by the normalisation of an APEC model in the following equation.

\begin{equation}
I \propto \frac{10^{-14}}{4\pi D_{A}^2}\int n_en_HdV
\label{eq.normapec}
\end{equation}
where $D_A$ is the distance to the GC (cm), 
and $n_e$ and $n_H$ are the electron and H densities (cm$^{-3}$), respectively. 
 We assume full ionisation with 10\% helium and 90\% hydrogen (n$_e$=1.2$n_p$).

Then the density (equation \ref{eq.density_3d}) is projected along the $z$ axis, and taking into account equation \ref{eq.normapec}, we go from a 3D distribution of the density to a 2D distribution of the intensity. 
The intensity is then transformed to flux using a count rate to flux conversion of cr2f = 1.376$\times10^{-11} \mathrm{erg\, cm^{-2}\,cts^{-1}}$ using PIMMs\footnote{https://cxc.harvard.edu/toolkit/pimms.jsp} and assuming a thermal plasma spectrum of temperature 7\,keV.
Finally, the 2D flux map model is fitted to our 2D excess Fe~XXV emission flux map. 
The best-fit parameters using both scalings of Model 2 (1:1, and [Fe]) are reported in Table \ref{tab.3d}.

\begin{table}
	\centering
	\caption{Flat power-law density model}
	\scalebox{0.9}{
		\begin{tabular}{@{}ccccc@{}}
			\hline  
		 scaling&$n_0$& $x_s$ & $y_s$ & $\beta$   \\
		 &\cmcube &arcmin &arcmin &  \\
		\hline
	[1:1]&0.115$\pm$ 0.001& $9.76\pm 0.04$ &$4.82\pm0.02$&  $0.388\pm0.001$ \\[3pt]
	[Fe]&0.109$\pm$ 0.001& $10.23\pm 0.08$ &$5.02\pm0.04$&  $0.411\pm0.001$ \\[3pt]
			\hline		
			\end{tabular}} 	
		\label{tab.3d}
	
		\end{table}

\subsection{Physical properties of the diffuse excess X-ray emission }\label{physicalproperties}

Assuming that all of the excess emission could be due to truly diffuse very hot plasma in the central degrees of the GC, we computed its physical properties using the best-fit model of the previous section (equation \ref{eq.density_3d}; Table \ref{tab.3d}). 

The thermal energy inside a sphere is given by: 
\begin{equation}
E_{\rm sphere}=2.64\times 10^{-8} kT\, R^{3}\, n_{H} \quad\quad \text{(erg)}.
\end{equation}
where $kT$ is the temperature in keV, R is the radius of the sphere in cm, and $n_{H}$ the hydrogen density in \cmcube{}. For a given volume element, and an arbitrary geometry of the plasma, the thermal energy density is:
\begin{equation}
U_{\rm th}=\frac{3\times2.64}{4\pi}\times 10^{-8} kT\, n_{H} \quad\quad \text{(erg \cmcube)}.
\label{eq.thermalenergy}
\end{equation}
We used equation \ref{eq.thermalenergy} and the best-fit model for the density in order to calculate the thermal energy contribution of each element in our 3D grid. 

From the modelling performed in the previous sections we find that the excess emission has an ellipsoidal shape and its bulk is concentrated within $\sim0.50^{\circ}$ (see Table \ref{tab.2d}).
We calculated the thermal energy, using this limit as the value of the semi-minor axis of an ellipsoid with major axis defined by the value of $q$ (see Table \ref{tab.2d}), and find a thermal energy of $E_{\rm th}\sim 2.5\times 10^{53}$\erg{} and $E_{\rm th}\sim 2.0\times 10^{53}$\erg{} for the 1:1 and Fe scaling respectively. When assuming higher helium abundances, smaller values of the thermal energy are obtained (i.e. for 15\% helium, $E_{\rm th}\sim 1.3\times 10^{53}$\erg{} for the 1:1 scaling). If the plasma consists only of helium then it could be confined in the GC (see the Introduction and \citealt{belmont05}).

We then measure the hot plasma sound speed using the formula:

\begin{equation}
c_s=\sqrt{\gamma\frac{kT}{\mu m_p}}
\end{equation}

\noindent where $k$ is the Boltzmann constant, $T$ the temperature of the gas, $\mu$ the mean molecular weight of the gas, $m_p$ the proton mass and $\gamma$ the adiabatic constant.
In our case we assume full ionisation with 10\% helium and 90\% hydrogen.
Therefore, the sound speed at 7\,keV is 1350 km\,$\rm{s^{-1}}$, which results in a sound crossing time of $t_s=0.6-1.3\times10^{5}$ yr, taking into account the different radii of the ellipsoid.
The power of the outflow is:

\begin{equation}
P=\frac{E_{\rm th}}{t_s},
\end{equation}

\noindent which we measure to be $P=0.8-1.5\times10^{41}$\ergs{} and $P=0.6-1.2\times10^{41}$\ergs{} for the 1:1 and Fe scaling respectively.

In the above calculations, in order to be conservative, we assume that the X-ray gas is volume-filling (filling factor $f=1$). This provides a lower limit for the measured density of the very hot gas.

\section{Discussion}\label{discussion}

\subsection{Excess Fe~XXV emission - comparison with previous works}

The percentage of the Fe~XXV line emission expected from stars, has been calculated in previous works using two  methods: scaling either SMDs or NIR maps to the 6.7\,keV X-ray emission towards the Galactic ridge.
In this section we present studies which have used either one or the other method and their most important findings.

\citet{uchiyama11}, used \suzaku{} observations and compared the Fe~XXV K$\alpha$ profile with an SMD model (see their appendix) made from NIR observations compiled by \citet{muno06} and using the results of \citet{launhardt02} and \citet{kent91}. They found a large Fe XXV X-ray emission excess compared to the scaled NIR luminosity, namely $19\pm6$ times larger for longitudes smaller than $l<0.2^{\circ}$ and $\sim$3.8 times larger up to $l=1.15^{\circ}$. They attributed this excess to either a new population of sources or an optically thin thermal plasma.

\citet{heard13sources} analysed \xmm\ data of the central 100$\times 100$ pc of the GC and compared the latitudinal and longitudinal profiles with the same SMD model used in \citet{uchiyama11}. They found an excess of the Fe~XXV emission compared to the NIR of a factor of $\sim$2. They attributed the different results compared to \citet{uchiyama11} to the different scaling chosen for the Galactic disc component of the SMD. Moreover, they explained the excess in terms of either a different kind of underlying source population, or an inaccurate SMD that does not account for all the mass enclosed in the NSC and/or the NSD. They also ruled out the possibility of a very hot, diffuse thermal plasma as the source of this excess since the properties of the X-ray spectrum match those of a larger number of IPs in the GC than other regions, a result that has been supported by \nustar{} observations towards the GC and bulge \citep{perez19}.

\citet{nishiyama13}, argued that the SMD models \citep{launhardt02} constructed from NIR maps, and used in the prior works of \citet{uchiyama11} and \citet{heard13sources}, could be subject to the influence of bright stars. For that reason they used NIR data to construct a stellar number density map \citep{yasui15}, which covers the central region of the Galaxy for $l$ within $\pm3.0^{\circ}$ and $b$ within $\pm 1.0^{\circ}$. They scaled this map (at $\ell>1.5^{\circ}$) with the longitudinal and latitudinal profiles of the 6.7\,keV line emission measured with \suzaku{} observations \citep{koyama07,uchiyama11}, and showed that the spatial distribution of the 6.7\,keV emission in the GC shows an excess of 50-80\% compared to the NIR distribution, thus favouring the diffuse, hot plasma scenario.

Our \xmm{} map is of much higher resolution and latitudinal coverage and allows us to safely scale with the SMD models at the \chandra{} deep region, where more than 80\% of the 6.7\,keV emission was attributed to unresolved point sources \citep{revnivtsev09}. 
The 6.7\,keV line emission excess we measure  using Model 1 ($\times$1.9 more than expected) is in agreement with the works of \citet{nishiyama13} and \citet{heard13sources}. However, our fiducial Model 2 results in a lower excess compared to what has been measured in all previous works ($\times1.3-1.5$ versus $>2.0$). 

\subsection{Origin of the excess Fe~XXV emission}

\subsubsection{The very hot gas explanation}
Regarding the physical origin of a truly diffuse hot plasma, many explanations have been proposed (see Introduction), which consist mainly of past star-forming activity or flaring of the central supermassive black hole, thereby thermalising the ISM.
Our calculations for the thermal energy ($E_{\rm th}= 2.0-2.5\times 10^{53}$\erg{}) in Sect.\,\ref{physicalproperties} are somewhat higher than the  estimates presented in \citet{uchiyama13} ($E_{\rm th}\sim 1\times 10^{53}$\erg{}). In addition, \citet{uchiyama13} estimate a density of the order of $\sim$0.05\,\cmcube{} while we measure a central density of $\sim$0.11\,\cmcube{} which drops as a function of radius.
\citet{ponti19} have found, for the chimneys,  a thermal energy of $E_{\rm th}=4\times10^{52}$\erg{}, with the power of the outflow being $P=4\times 10^{39}$\ergs{}. Our measurements for the central $\sim$2$^{\circ}$ of the GC are more than one order of magnitude higher. Therefore, if a truly diffuse, very hot plasma is present, it could possibly power the chimneys, or whatever the hot plasma source is, it could power both the chimneys and the very hot plasma.

The patchy shape of our X-ray profiles (see Figs.\,\ref{fig.profilelat} and \ref{fig.profilelong}) with bumps coinciding with known SNRs and super-bubbles, favours the star-forming scenario.\citet{uchiyama13} calculated that if all thermal energy observed is produced solely by SNe, a rate of $\mathrm{>5\times 10^{-3}\,yr^{-1}}$ is necessary. They deem this value unreasonably high, given the mass measurements of the GC region.
Using the revised total thermal energy of the hot plasma we calculated in Sect.\,\ref{physicalproperties}, we find a requisite SN rate of $\mathrm{>1.9\times 10^{-3}\,yr^{-1}}$. Estimations of the SN rate in the central degrees of the GC, mostly in the CMZ region, yield values of  $\mathrm{0.2-1.5\times 10^{-3}\,yr^{-1}}$ \citep[][]{crocker11,ponti15}. 
Assuming that most of the energy released by SN is converted into thermal energy of the hot plasma (the efficiency of energy transfer from supernova blast waves to the ISM is much higher if the explosion takes place within a pre-existing superbubble, as we have here), the observed SN rate is nearly sufficient to supply the requisite energy. Thus we conclude that SNe might be providing a substantial portion of the energy needed to heat the plasma in the GC.

Regarding the progenitor type of the SNRs responsible for the 6.7\,keV emission, both Type Ia and core collapse SN could contribute to the Fe~XXV emission. However, Type Ia SN typically have an Fe\,K$\alpha$ centroid below 6550\,eV, whereas core collapse SN produce lines with a higher centroid energy, more consistent with the emission we get from the GC \citep[see][and their figure 1]{Yamaguchi14}. Therefore, we expect that core collapse SN will make the largest contribution to the 6.7\,keV line we observe in the GC. 
However, it is important to note that, for the 6.7\,keV emission line to be detectable, SNRs would need to be neither too evolved nor too young \citep[][]{Yamaguchi14}. This places more constraints in order to produce the necessary high temperature we observe, suggesting that SNe explosions might not be the only source of a possible very hot plasma.

Moreover, we have checked that the X-ray excess profile does not follow well the dense gas mass profile. Indeed, it is well known that the gas distribution in the CMZ (R<300pc) is highly asymmetric, with roughly 3/4 of gas being at positive longitudes and only 1/4 at negative longitudes \citep[e.g.][]{henshaw22}, while in contrast the X-ray profile is roughly symmetric. So the two profiles are qualitatively different. However, we note that this does not exclude the possibility that some of the observed X-ray excess is due to a hot plasma originating in supernova explosions. Although dense gas correlates with ongoing star formation, we do not expect it to correlate with the location of supernova explosions. By the time the first supernova explosions occur (>4 Myr after the formation of a star), the stars have already decoupled from the gas cloud in which they were born (the decoupling between gas and stars occurs on timescales of a few Myr, as discussed, for example, in section 3.2 of \citealt{sormani20b}).

Another source of energy to produce a very hot diffuse plasma component could be energy released from past flaring activity of the supermassive black hole, Sgr\,A$^{*}$. 
Signs of dramatic recent flaring are evident by their X-ray `echoes' across the CMZ, with the most recent one having occurred about 120 years ago, with total emitting power of $\sim$$10^{47}$erg and a hard X-ray spectrum \citep[][]{sunyaev93,koyama96,ponti13, churazov17a}.
Therefore, too many episodes of dramatic flaring of similar energy ($\sim$ 1 episode/yr) during the past $\mathrm{0.5-1\times10^5\,yr}$ would be required in order to explain the energetics we measure today and no such episode has been observed in the last 30-40 years.

\subsubsection{Sources with higher metallicity in the GC}\label{highmetallicitysources}

The measurements of the iron abundance of the underlying stellar populations in the GC \citep{schulteis21,fritz21} point to a  metallicity difference between the NSD and the bulge/bar of  a factor of $1.35$ which translates into a multiplicative flux difference of $1.25$ (see Sect.\,\ref{metallicityimplications}). 
Moreover the existence of high metallicity sources (about twice solar) in the NSC and NSD is further supported by near infrared studies \citep{feldmeierkrause17,nogueraslara20,schoedel20}.

Since an enhanced iron abundance source population exists in the GC, we would expect to see the impact of higher metallicities also in the X-rays (i.e. higher iron abundances and consequently EWs of the 6.7\,keV line). \citet{uchiyama13}, after fitting \suzaku{} spectra from the GC and ridge, as a result of the overall high abundances in the GC, consider that GC X-ray emission would require a different type of point source population (i.e. with higher temperature and abundances) than the one in the Galactic ridge, which they deem artificial. \citet{yamauchi16} studied iron line EWs of the Galactic diffuse emission and found a 1.19 and 1.23 increase of the EW of the 6.7\,keV line in the GC compared to that of the bulge and ridge respectively, in line with the values found by \citet{schulteis21} and \citet{fritz21}.

\begin{figure}
 	\centering
 		\includegraphics[width=1.0\columnwidth]{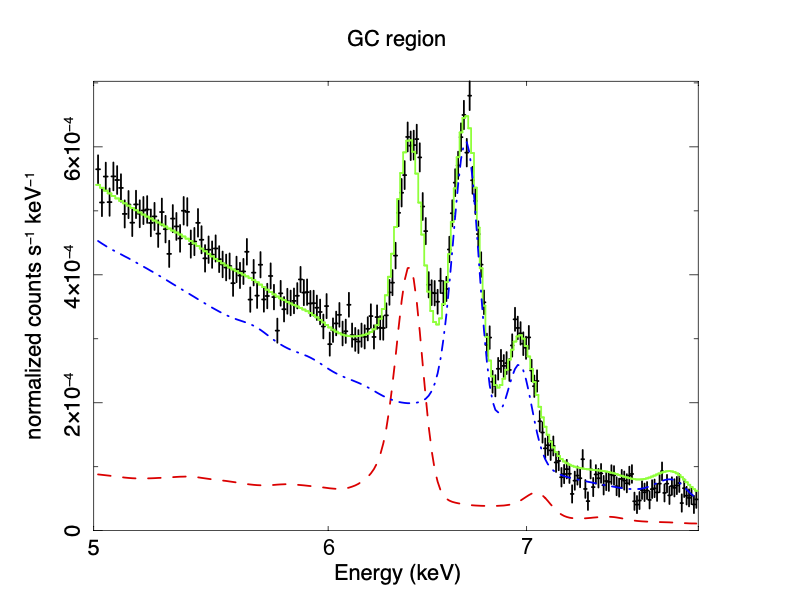}
 		\caption{\xmm{} spectrum of the GC region. Combined MOS spectrum (black crosses) of the GC region fitted with a reflection model (dashed-red line), and a thermal plasma model (dash-dotted blue line).}
 		\label{fig.spectra}
 
 \end{figure}

In order for all the excess to be explained by metallicity differences, as calculated in Sect.\,\ref{metallicityimplications}, a $\sim1.9$ higher metallicity would be required for the NSD in comparison to the bulge/bar.
To test further whether a 1.9 times higher scaling for the NSD could be justified, we extracted \textit{XMM-Newton} EPIC MOS spectra from two circular regions of size 12 arcmin, one close to the GC ($\ell=359.53$, $b=0.04$; hereafter GC region) and one close to the \chandra{} deep region ($\ell=0.04$, $b=-1.53$; hereafter the scale region) after excising all the bright sources.
The combined MOS spectra show emission lines at 6.4\,keV due to reflection, and at 6.7\,keV, and 6.9\,keV due to hot plasma emission or emission coming from unresolved compact objects (see Fig.\,\ref{fig.spectra} for the GC region).
We fitted the combined MOS spectra of the two regions using a phenomenological model (\texttt{po+gaussian+gaussian+gaussian}) to describe the continuum and the Gaussian lines. We measured the EWs of the 6.7\,keV line (GC region: $\sim$ 255\,eV; Scale region: $\sim$215\,eV), that give a ratio of $\sim1.2$ for the EW$\mathrm{_{6.7}}$ of the GC over the scale region, which is in agreement with previous works.
We also tried a physically motivated model (\texttt{crefl16+vapec}), where we used a reflection model template \citep[see][]{churazov17} to represent the reflection emission from molecular gas in the GC, and a thermal component with variable abundances to represent the emission lines at 6.7\,keV, and 6.9\,keV. We then left the iron abundance free to vary. Our model fits yield $\mathrm{[Fe]=0.67_{-0.05}^{+0.03}}$ and $\mathrm{[Fe]=0.36_{-0.10}^{+0.13}}$ for the GC and the scale region respectively, giving a ratio for the two regions of  $\mathrm{GC_{[Fe]}/scale_{[Fe]}=1.86}$.
In Fig.\,\ref{fig.spectra} we show the extracted spectrum for the GC region and the best-fit model using thermal emission (vAPEC) plus reflected emission \citep[][]{churazov17}.
Therefore, we see that using different models a higher ratio close to 1.9 is possible. Of course this difference could originate from the existence of hot plasma in the GC but nevertheless the factor 1.9 needed to explain almost of all the excess due to higher metallicity of the sources in the GC can be reproduced.

\begin{figure}
 	\centering
 		\includegraphics[width=1.0\columnwidth]{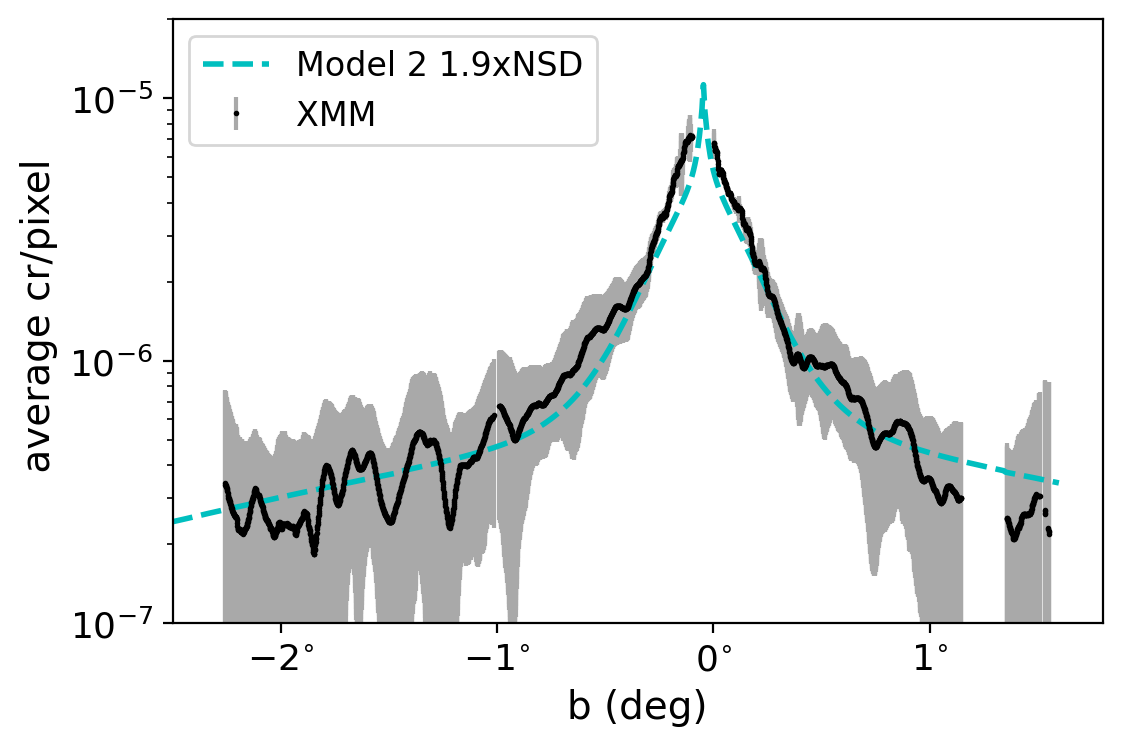}
 		\includegraphics[width=1.0\columnwidth]{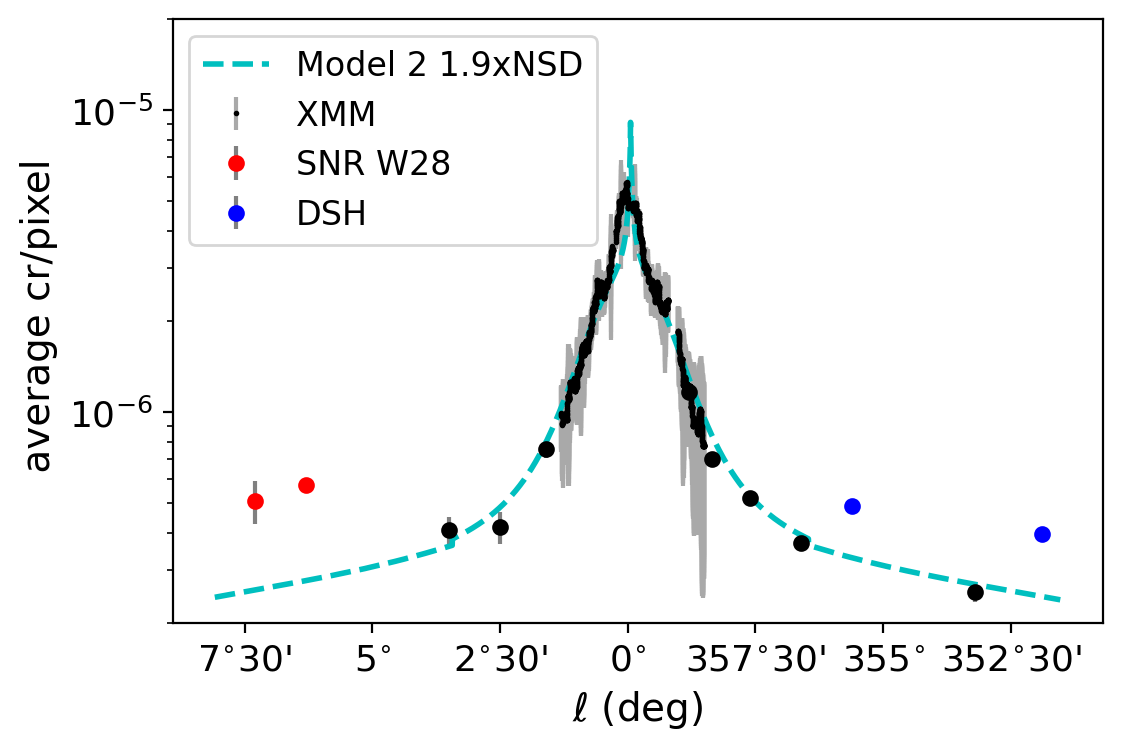}
 		\includegraphics[width=1.0\columnwidth]{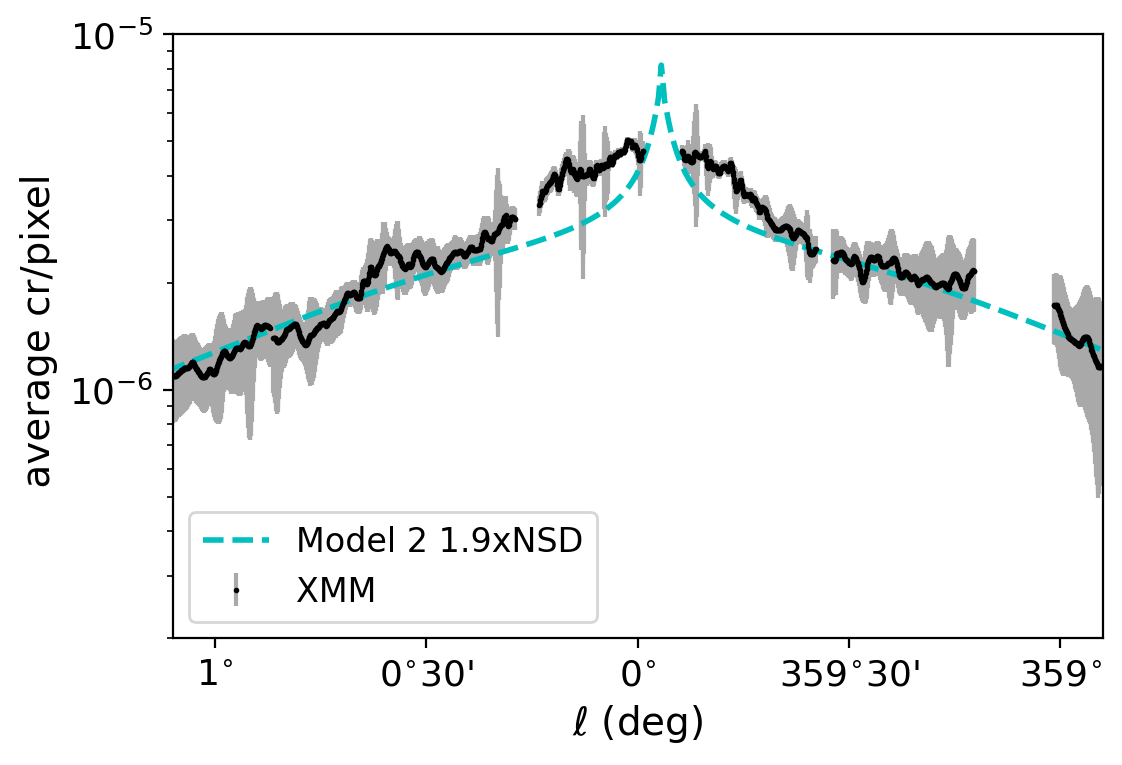}
 		\caption{Scaled SMD profiles assuming a higher NSD iron abundance compared to that of the bulge. Latitudinal (top panel) and longitudinal (middle panel) profiles of 6.7\,keV emission along with the scaled Model 2 assuming that the NSD metallicity is 1.9 times higher than that of the bulge. Bottom panel is the same as middle panel, only within $\ell~=~\pm 1^{\circ}$.}
 		\label{fig.profilelatscaledNSD}
 
 \end{figure}

We show in Fig.\,\ref{fig.profilelatscaledNSD} how the latitudinal and longitudinal profiles presented in Figs.\,\ref{fig.profilelat} and \ref{fig.profilelong} change when we scale the NSD model component by 1.9 times compared to the bar/bulge. We see a very good agreement between the X-ray profile and the scaled Model 2 with no excess X-ray emission remaining for the latitudinal profile while for the longitudinal profile (see Fig.\,\ref{fig.profilelatscaledNSD} bottom panel) a small excess is visible in the location of the SNR G0.61+0.01, and a bit larger excess in the very central $\ell\sim\pm0.3^{\circ}$ of the GC. 
In fact the remaining excess (1.15 times more than expected by the SMD) matches the width of the X-ray chimneys \citep[][]{ponti19}, which could indeed be explained by very hot plasma in the GC due to the star-forming activity, and/or past flares of Sgr\,A$^{*}$ thermalising the ISM. 
We calculate the thermal energy of the remaining excess X-ray emission located in the central $\ell\sim\pm0.3^{\circ}$ and $b\sim\pm0.15^{\circ}$. We find a value of $E_{\rm th}\sim 2.0\times 10^{52}$\erg{} for its thermal energy, with a power of $P\sim4.0\times10^{40}$\ergs{}, with the latter being an order of magnitude higher than that of the X-rays chimneys.
If all the thermal energy is produced by SNe, then a rate of  $\mathrm{>0.6\times 10^{-3}\,yr^{-1}}$ would be required. This is in agreement with measurements of the SN rate in the CMZ region \citep[$\mathrm{0.2-1.5\times 10^{-3}\,yr^{-1}}$][]{crocker11,ponti15}.

\subsubsection{A new population of sources}

The agreement between the X-ray profiles and the scaled Model 2 for 1.9 times higher NSD metallicity compared to that of the bar/bulge is quite remarkable (Fig.\,\ref{fig.profilelatscaledNSD}). The stellar distribution follows the X-ray emission so well that it seems quite fortuitous and therefore improbable for the very hot plasma to be the main reason for the excess measured initially. The fact that the iron abundance enhancement can be inferred from the corresponding X-ray spectra of the GC and bar/bulge (see previous section) points to the existence of a new population of point sources in the GC. This higher metallicity point source population, along with a moderate amount of truly diffuse hot plasma in the central half degree of the GC, can explain all the previously unexplained excess.

This new population of sources should be present in the GC and should exhibit stronger (a higher EW) 6.7\,keV line emission, which could have been enriched by past star-forming activity and SNe.
This could explain also why past studies \citep[e.g.][]{uchiyama13,yamauchi16,nobukawa16} have failed to reproduce the observed equivalent widths (EWs) of the 6.7\,keV line (together with the observed 6.4 and 6.9 keV lines) resulting from global fits of the GC region with a combination of known sources such as magnetic CVs, non-magnetic CVs and ABs.

In the CMZ region, a handful of sources which show very strong 6.7\,keV line EWs (>1 keV) and fall in the category of very faint X-ray transients has been identified. Their exact nature is not yet known \citep[e.g.][]{sakano05,hyodo08}.
If this population of sources extends to lower luminosities, then it could justify an enhanced iron abundance in the GC. This would require a rather large, faint population, and be present only in the central degree of our Galaxy. Results towards this direction have been presented by \citet{zhu18}. They fitted cumulative spectra of point sources in the GC (mainly the NSC) and the Chandra deep field \citep[][]{revnivtsev09} and found that the faint population of sources in the GC can have as much as $\sim$4.5 times higher  6.7\,keV EWs than those in the bar/bulge.
Further exploration of the 6.7\,keV EWs also for the NSD faint point source population would allow safe conclusions to be drawn on the metallicity differences between the GC and the Bar/Bulge.

\subsection{Sources of uncertainty}

Throughout this work a smaller value of the X-ray excess 
 was measured compared to previous works, regardless of the chosen scaling of the SMD to the X-ray data, namely the 1:1 scaling, the [Fe] scaling and the 1.9 higher metallicity scaling. 
There are various factors that contribute to the uncertainty of these calculations, which make it difficult to give a unique estimate of the uncertainty. Therefore, in this section we summarise  all possible sources of uncertainty mentioned throughout this work.

\textit{Uncertainties introduced by the X-ray data:} For the \xmm{} data presented in this work, the uncertainties on the profiles are based on the errors introduced in the count and background images of the EPIC detectors (see Sect.\,\ref{xraymosaic}). Observations closer to the GC result in having much smaller errors (averaged per pixel; see e.g. Fig.\,\ref{fig.profilelat}). For example observations within $\pm$0.25 degrees from the GC have on average error less than 3\%, while between 0.5 and 1 degree for the GC can reach up to 30\%. This is the result of many more \xmm{} pointings dedicated to the coverage of the CMZ and the  Sgr\,A$^{*}$ regions (see Table \ref{tab.xmm}).

\textit{Uncertainties connected to the scale region:} The scale region that was used (see region within red dotted lines in Fig.\,\ref{fig.profilelat} and Sect.\,\ref{profiles}) to scale the SMD to the X-ray data has uncertainties introduced by the X-ray data of the order of $\sim$5\%. This more extended region was used in order to minimise the uncertainties that are quite large ($>$50\%) per \xmm{} pixel for latitudes close to the \chandra{} deep region.

\textit{Uncertainties introduced by the SMD:} Each component (NSC, NSD, Bar/bulge, and disc) of the SMD contributes with its own uncertainties. However, for our calculations the most important contributions, since we are concentrating on the central 3 degrees of the GC, are the NSD and the Bar/Bulge components (see Sect.\,\ref{SMDs}).
The uncertainties introduced by the NSD and bar/bulge components of our fiducial Model 2 are of the order of $\sim$10\% \citep{sormani22nsd,sormani22bar}.
These uncertainties depend on the shape of the density profile, which is difficult to estimate due to high extinction in the GC and on the uncertainties on the metallicities of stars \citep{schulteis21}. Therefore, the uncertainties of the NSD and bar/bulge components give a lower bound to the overall uncertainty originating from the SMD.

Overall we would expect in the central half degree an uncertainty of the order of $\sim$10\%. Although this should be treated as a lower estimate since there are many factors that contribute to the uncertainty and are connected to the complexity of the environment near the GC (see discussion before). We nevertheless see it as a concrete possibility that given all these factors the excess could reduce almost to zero in the future with more accurate measurements.

\subsection{Possible connection to Fermi-LAT Galactic centre excess}

In addition to the hard X-ray excess (6.7\,keV) in the GC, a \textit{Fermi}-LAT $\gamma$-ray excess has been identified, after removal of  point sources, between 1 and 3 GeV \citep[see review and references therein;][]{murgia20}.
 The main explanations for this excess that have been considered are: 1) annihilating dark matter, 2) unresolved point sources such as millisecond pulsars, 3) cosmic ray (CR) outbursts at the GC that could be originating from past activity of Sgr\,A$^{*}$ or starburst events (with the Fermi bubbles as clear evidence of the past activity), and 4) an enhancement of  CR source populations or of the intensity of the interstellar radiation field. The first two of these explanations are the most widely considered. 
The spatial morphology of the $\gamma$-ray emission is considered to be consistent with being spherical, it is brightest towards the GC and extends up to $b=10^{\circ}$ \citep[see figure 2 of][]{murgia20}. 
The spherical morphology does not favour an origin from CR sources or CR outbursts since it should broadly trace the distribution of the molecular gas in the GC which is highly flattened. However, other morphologies have been suggested such as boxy or X-shaped following the stellar distribution in the Galactic bar/bulge \citep{macias18,bartels18,macias19}.

\begin{figure}
 	\centering
 		\includegraphics[width=1.0\columnwidth]{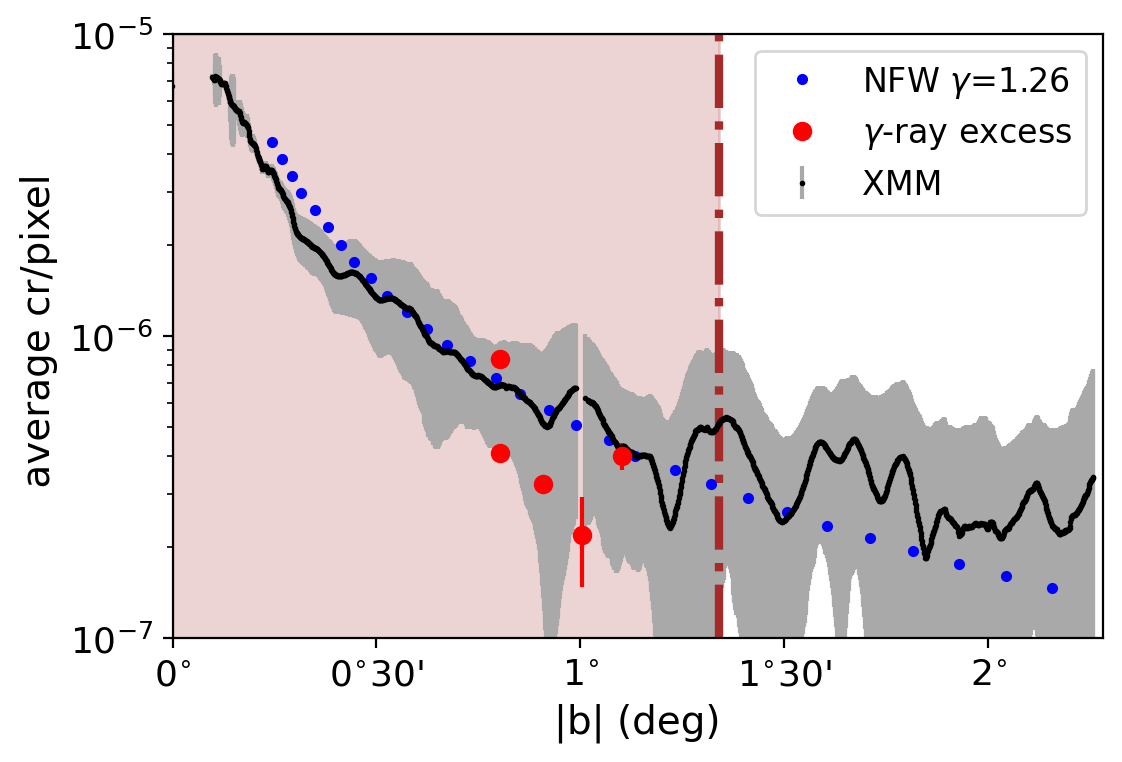}
 		\caption{Comparison of the X-ray and the $\gamma$-ray excesses in the GC. \xmm{} Fe~XXV profile (black line) adopted from Fig.\,\ref{fig.profilelat} for negative latitudes. With red points we show the GC excess intensity at 2 GeV as measured by various studies \citep{hooper11,boyarsky11,gordon13,abazajian14,daylan16}. The blue dotted line represents the emission from dark matter annihilation for a Navarro–Frenk–White distribution \citep[NFW][]{navarro97}, which best represents the $\gamma$-ray excess data. Both red points and blue points associated with the $\gamma$-ray excess are normalised along the y-axis to match the X-ray profile for illustration purposes. 
 		For more details see \citet{murgia20} and their figure 2.
 		 }
 		\label{fig.fermi}
 \end{figure}

It would be interesting to examine whether a common physical origin of the X-ray and $\gamma$-ray excess could be possible. One of the main differences between the two excesses is their extent. The X-ray excess (when we assume 1:1 scaling) is no longer visible at latitudes above $b=\pm 1.2^{\circ}$ since at $b=-1.4^{\circ}$ deep \chandra{} observations have resolved almost all of the diffuse emission into point sources \citep[][]{revnivtsev09}, while the $\gamma$-ray excess extends up to  $b=\pm10^{\circ}$. 
If we compare the \xmm{} latitudinal profile (up to 2$^{\circ}$), with point sources removed but no unresolved emission removed (see Fig.\,\ref{fig.profilelat}), to the equivalent distribution for $\gamma$-ray  \citep[figure 2 of][]{murgia20} we find that the slopes of the two distributions are consistent with being the same, given the large dispersion of the $gamma$-ray excess points (red points in Fig.\,\ref{fig.fermi}), and the similarity to the slope of their best-fit model (blue points in Fig.\,\ref{fig.fermi}). This similarity could  indicate a common origin for the two excesses in the central degree of our Galaxy. 

Within $b\sim\pm1.3^{\circ}$ the gas-related $\gamma$-ray emission from the inner Galaxy is significant \citep[vertical line in figure 2 of][and brown vertical line in Fig.\,\ref{fig.fermi}]{murgia20}.
Indeed in the inner 2 kpc of our Galaxy most of the emission (90-95\%) in $\gamma$-rays originates from interactions of CRs with the interstellar medium \citep[][and references therein]{murgia20}. This process needs to be modelled in detail, and uncertainties connected with its modelling can affect significantly the inferred properties of the excess emission. Therefore, a different population of CR sources, or the intensity of the interstellar radiation field could alter the interpretation of the $\gamma$-ray excess observed.
One of the candidate classes of sources for Galactic CRs is SNRs \citep[e.g.][]{helder12}, and indeed they are abundant in the central degree of our Galaxy. Also SNRs are among the candidates of 6.7\,keV excess emission. Therefore, an overabundance of SNRs that are sources of both CRs and 6.7\,keV emission could provide a common explanation of the two excesses, at least in the inner 2$^{\circ}$.
However, that would require that a different mechanism be responsible for the $\gamma$-ray excess at higher latitudes. 
On the other hand, if high-metallicity source populations are responsible for almost all the excess we see in X-rays from the GC (see Sect.\,\ref{highmetallicitysources}), then the similarity of the profiles of the 6.7 keV line and the gamma-ray flux suggests that old stellar-mass objects (plausibly a different sub-class) are responsible for the gamma-ray excess. The much lower resolution of \textit{Fermi}-LAT compared to that of \xmm{} and other X-ray telescopes could then explain the difference in the extent of the two excesses since a larger number of point sources are presumably unresolved in $\gamma$-rays. The point source origin scenario is further supported by studies that point to the \textit{Fermi}-LAT GeV excess being the tracer of stellar mass (i.e. unresolved faint source population) in the Galactic bulge since some investigators have found evidence for a distinct gamma-ray source that is traced by the ``nuclear bulge”
 \citep[e.g.][]{macias18,bartels18,macias19}.

\section{Summary}\label{summary}

In this work we have analysed all (370) available \xmm{} observations ($\sim$6.5\,Ms) of the GC and disc spanning  the region out to $l=\pm 10^{\circ}$ and $b=\pm 2.0^{\circ}$, in order to study the 6.7\,keV line emission and physical properties of the emitting plasma, mainly in the central degrees of the GC. We are able for the first time to scale the SMD models using the \chandra{} deep region where \citet{revnivtsev09} found that more than 80\% of the emission is produced by unresolved X-ray point sources. We find that:

\begin{itemize}
    \item When we subtract the point source contribution using the SMD models, there remains an excess of 6.7\,keV emission that is $\sim$1.3 to 1.5 times larger than what is predicted by the SMD model and is concentrated in the central $2^{\circ}\times 0.5^\circ$. The excess we find is lower than the one found in previous works which is the result of our use of a different and more recent SMD, and with its scaling to the X-ray emission accounting also for metallicity differences between the NSD and bar/bulge stellar populations.  
    \item The shape of the longitudinal profile shows enhanced emission at the locations of known SNRs, pointing to the contribution of past star-forming activity for at least a portion of the excess emission.
   
    \item The thermal energy ($\sim2\times10^{53}$ erg) and the power ($0.6-1.2\times10^{41}$ erg\,s$^{-1}$) of the implied outflow we calculate in case this excess is due to hot plasma, are high enough to power the outflows  we observe in the GC (i.e. the X-ray chimneys). However, SNe or dramatic flares from the supermassive black-hole (as constrained by X-ray echos)  alone are apparently inadequate to reproduce these values.
    
    \item Almost the entire X-ray excess can be explained by assuming an iron abundance of $\sim1.9$ times higher for the stellar populations in the NSD compared to those in the bar/bulge. We were able to reproduce this value by fitting spectra from these two regions.

    \item With the $\sim$1.9 times scaling of the NSD, the X-ray profile and SMD show a very good agreement, with X-ray excesses remaining within the region of a known SNR, and in the central $\ell\sim\pm0.3^{\circ}$ and $b\sim\pm0.15^{\circ}$ of the GC. The remaining excess in the GC has a longitudinal width similar to that of the X-ray chimneys, and a  thermal energy of $\sim2\times10^{52}$ erg which can be reproduced by the estimated SNe rate in the GC.
 
\end{itemize}

Overall the above point to the existence of a higher iron metallicity source population in the GC. Such a population, along with a moderate amount of truly diffuse very hot plasma in the central half degree of the GC can explain the hard X-ray emission we observe.

\begin{acknowledgements}
This project acknowledges funding from the European
Research Council (ERC) under the European Union’s Horizon
2020 research and innovation programme Hot Milk (grant agreement No
865637). KA thanks Dimitrios L. Anastassopoulos for the helpful discussions.
MM acknowledges support from NASA under grant GO1-22138X to UCLA. MSC acknowledges support from the Deutsche Forschungsgemeinschaft (DFG) via the Collaborative Research Center (SFB 881, Project-ID 138713538) ``The Milky Way System'' (sub-projects A1, B1, B2 and B8) and from the European Research Council in the ERC Synergy Grant ``ECOGAL - Understanding our Galactic ecosystem: From the disc of the Milky Way to the formation sites of stars and planets'' (project ID 855130).
MS acknowledges support from the Deutsche Forschungsgemeinschaft through the grants SA 2131/13-1, SA 2131/14-1, and SA 2131/15-1.
IK acknowledges support by the COMPLEX project from the European
Research Council (ERC) under
the European Union’s Horizon 2020 research and innovation programme grant
agreement ERC-2019-AdG 882679.
EDT acknowledges support by the MW-WINDS project from the European Research Council (ERC) under grant agreement no. 101040751. We also made use of NASA’s Astrophysics Data System Bibliographic Services. 

\end{acknowledgements}



\bibliographystyle{aa} 
\bibliography{references} 


\begin{appendix}

\section{Acronyms and scaling of MOS exposure maps}

In order to help the reader better follow the acronyms used throughout this work, we present them in alphabetical order in Table \ref{tab.acronyms}. 

\begin{table}[!h]
	\centering
	\caption{Acronyms used throughout this work}
		\begin{tabular}{@{}ll@{}}
			\hline  
			Acronym & full name \\
			
			 \hline
	AB& active binary \\
	CV &  cataclysmic variable \\
	CMZ& central molecular zone \\ 
	DSH& dust scattering halo\\ 
	EA & effective area \\
	EM & exposure map \\
	GC  & Galactic centre\\
    IP & intermediate polar\\
    ISM & interstellar medium\\
    NIR & near-infrared\\
    NSC & nuclear stellar cluster\\
    NSD & nuclear stellar disc\\
    SMD &  stellar mass distribution\\
    SNR & supernova remnant \\ 
    XRB &  X-ray binary \\
			\hline			
			\end{tabular} 	
		\label{tab.acronyms}
		\end{table}

Moreover, in this section we describe how exactly the MOS exposure maps (EMs) were scaled in order to account for the differences in the effective area (EA) between MOS and pn detectors.
At the energy range of 6.62-6.8 keV the effective area of the PN detector is about 4 times larger compared to that of the MOS detectors (see XMM handbook), and this should be taken into account when one creates a map including all detectors.
Assuming a thermal spectrum of kT=7.0 keV from PIMMs tool, depending on the different filters of \xmm{} we get values of $EA_{pn}/EA_{MOS}=0.21-0.24$.
Moreover, we calculated the ratio of the pn over the MOS effective area, looking at an RMF file extracted near the GC in the 6.62 to 6.8 keV band. We find a mean value of $EA_{pn}/EA_{MOS}\sim0.22$.
We then multiplied the MOS exposure maps by the factor $EA_{pn}/EA_{MOS}=0.22$. The all-detector exposure map is then defined as $EM=EM_{pn}+0.22\times(EM_{MOS1}+EM_{MOS2})$. The resulting count rate is then representative of the pn detector, and can be transformed to flux using the equivalent count rate to flux conversion factor. 

\begin{figure}[!h]
 	\centering	\includegraphics[scale=0.5]{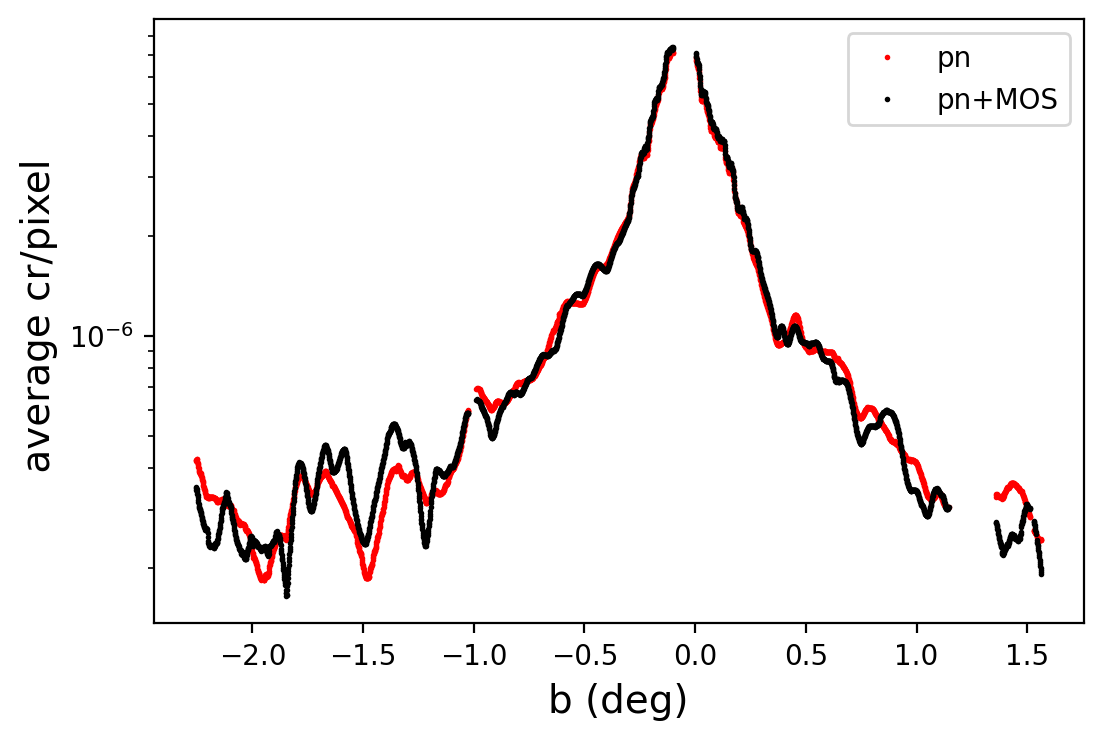}
 		\caption{Comparison between \xmm{} EPIC-pn and pn+MOS latitudinal profiles. Average count rate per pixel over Galactic latitude ($b$), extracted from a profile of 0.5 deg width centred at Sgr\,A$^{*}$. With red dots we show the profile extracted from the EPIC-pn mosaic, while with black dots we show the profile extracted from the pn+MOS mosaic, after scaling the MOS exposure maps to account for the effective area differences between MOS and pn at the 6.62-6.8 keV band. 
 		}
 		\label{fig.factor}
 \end{figure}

To examine the robustness of this calculation we extracted two profiles of width 0.5$^{\circ}$ (centred at Sgr\,A$^{*}$) along the Galactic latitude and longitude of the \xmm{} map (Fig.\,\ref{fig.factor}), using only the pn detector (red dots), and the pn+MOS detectors (black dots) where the exposure maps of the MOS detectors are weighted according to the procedure described. We see that the 0.22 factor is representative of the differences in the effective areas  between the instruments, since they agree very well. Moreover, we notice that the pn+MOS profiles show less noise at certain areas than the pn profiles.

 \FloatBarrier

\section{\xmm{} observations}

In Table \ref{tab.xmm} we present all \xmm{} observations used in this paper except those already presented in \citet{ponti15}. We report the OBSID number, the total and clean exposure time for each camera (in seconds), as well as the custom cut-offs (in counts per second) used for the background filtering after visual inspection of the background light curves. For observations where the threshold column is empty, the standard values of 2.5 cts\,s$^{-1}$ and 8.0 cts\,s$^{-1}$ for the EPIC MOS and EPIC pn exposures were used respectively.
We use the headings `CMZ', `GCL', `Sgr\,A$^{*}$', and `Heritage' for the scans of the Central molecular zone, the Galactic centre Lobe,  Sgr\,A$^{*}$ and the inner Galactic disc respectively. For all other observations we use the heading `Serendipitous'.

\clearpage
\onecolumn
\begin{longtable}{lccccccccc}
\caption{\label{tab.xmm}\xmm{} observations}\\
\hline
\hline
OBSID& Exp pn  &  Exp M1  &  Exp M2 & Exp pn  &  Exp M1  &  Exp M2 &&Threshold&   \\
&sec&sec & sec& sec &sec &sec &pn (cts\,s$^{-1}$) &M1 (cts\,s$^{-1}$)& M2 (cts\,s$^{-1}$)\\
\hline
\endfirsthead
\caption{continued.}\\
\hline
\hline
OBSID& Exp pn  &  Exp M1  &  Exp M2 & Exp pn  &  Exp M1  &  Exp M2 &&Threshold&   \\
&sec&sec & sec& sec &sec &sec &pn (cts\,s$^{-1}$) &M1 (cts\,s$^{-1}$)& M2 (cts\,s$^{-1}$)\\
\hline
\endhead
\hline
\endfoot
\hline
	&&&&\textit{Serendipitous}&&&&& \\
0001730201   &   22597   &  25761   &  25733   &  13397   &  19472   &  19407   & 5.0 &  1.8 &  1.8  \\       
0030540201    &   0       &  3834    &  3834    &  0       &  3834    &  3834    & & & \\      
0050940201    &   19787   &  23785   &  23780   &  13528   &  22051   &  22519   & & & \\      
0085580201    &   7148    &  9142    &  9142    &  4549    &  8783    &  8990    & & & \\      
0085580501    &   5000    &  6942    &  6942    &  5000    &  6942    &  6942    & & & \\      
0085581001    &   5335    &  7925    &  7925    &  5335    &  7925    &  7925    & & & \\      
0085581101    &   5000    &  6693    &  6693    &  2400    &  5088    &  5245    & & & \\      
0085581601    &   0       &  4604    &  4379    &  0       &  4604    &  4379    & & & \\      
0099760201    &   47578   &  49557   &  49518   &  35816   &  39794   &  39578   & 4.0 &  1.5 &  1.5  \\      
0103261201     &   4397    &  6677    &  6669    &  4300    &  6677    &  6669   & 6.0 &      &      \\     
0103261301     &   5000    &  7626    &  7626    &  5000    &  7626    &  7626   & & &  \\     
0135742201     &   5000    &  6617    &  6617    &  4500    &  6565    &  6565   & & &  \\     
0135742301     &   5000    &  6617    &  6617    &  3200    &  3760    &  3916   & 6.0 &  1.5 &  1.5   \\     
0135742401     &   4948    &  6552    &  6555    &  2648    &  4696    &  4794   & 6.0 &  1.8 &  1.8  \\     
0135742601     &   5883    &  5616    &  5595    &  3852    &  4911    &  4917   &&&  \\     
0135742801     &   5000    &  6617    &  6617    &  4300    &  5837    &  5889  & 6.0 &  1.8 &  1.8   \\     
0135742901     &   5000    &  6617    &  6617    &  3600    &  5213    &  5525  & 5.0 &  1.5 &  1.5  \\     
0135743101     &   4999    &  6617    &  6622    &  4499    &  6617    &  6622    &&& \\     
0135960101     &   18010   &  20621   &  20621   &  12810   &  15909   &  16377  & 6.0 &  1.5 &     \\     
0144271401     &   5326    &  6955    &  6960    &  3926    &  6643    &  6960    &&& \\     
0145840101     &   45562   &  48433   &  48454   &  13311   &  23725   &  24095 & 4.0 &  1.5 &  1.5    \\     
0145840201     &   33841   &  33946   &  33897   &  14901   &  25615   &  26018  & 5.0 &  2.0 &  2.0  \\     
0145970101     &   19019   &  21559   &  21567   &  19019   &  21559   &  21567   &&& \\     
0145970401     &   30829   &  32467   &  32469   &  17831   &  26331   &  26905 & 4.0 &  1.5 &  1.5   \\     
0148090501     &   9246    &  10949   &  10970   &  5336    &  8364    &  8526 & 4.0 &  1.5 &  1.5    \\     
0150220101     &   19879   &  30389   &  30366   &  19879   &  30389   &  30366  & & &  \\     
0152832801     &   7781    &  9399    &  9407    &  7581    &  9399    &  9407   & & &  \\     
0154750301     &   0       &  34799   &  35069   &  0       &  20124   &  19968  & & &  \\     
0164561101     &   18469   &  18400   &  18406   &  5172    &  17880   &  17886  & 6.0 &  1.6 &  1.6  \\     
0164561401     &   32475   &  32407   &  32667   &  13576   &  29737   &  31172  & 6.0 &  1.5 &  1.5  \\     
0201200101     &   15842   &  17315   &  17315   &  3698    &  16336   &  16752   &&& \\     
0205350101     &   51408   &  51590   &  51619   &  30759   &  39673   &  40447 &     &  1.5 &  1.5    \\     
0206590201     &   18968   &  20595   &  20590   &  17068   &  20595   &  20590  & & &  \\     
0206990101     &   8222    &  14295   &  14298   &  7922    &  14295   &  14298  & & &  \\     
0301730101     &   59848   &  61460   &  61480   &  59848   &  61460   &  61480  & & &  \\     
0301881001     &   14143   &  16019   &  14815   &  14143   &  15811   &  14815 &     &  2.0 &    \\     
0302570101     &   27208   &  31123   &  31135   &  27109   &  31123   &  31135  &&&  \\     
0307110101     &   27585   &  31538   &  31536   &  13041   &  23128   &  23907  &&&  \\     
0402040101     &   33505   &  35071   &  35087   &  18013   &  34984   &  35087  &&&  \\     
0402280101     &   42195   &  43759   &  43777   &  40606   &  43530   &  43576  &&&  \\
0405640201     &   20665   &  22621   &  22607   &  20466   &  22052   &  22318   &     &  2.0 &  2.0 \\     
0405640401     &   10207   &  12156   &  12161   &  5007    &  9351    &  9773    & 6.0 &  2.0 &  2.0 \\     
0405640501     &   9672    &  11618   &  11626   &  4472    &  8816    &  9290    & 6.0 &  2.0 &  2.0 \\     
0405640601     &   12066   &  13619   &  12320   &  10172   &  13619   &  12320  & 6.0 &      &      \\     
0405640801     &   13301   &  15247   &  15244   &  9503    &  13551   &  14129   & 6.0 &  2.0 &  2.0 \\     
0405640901     &   12926   &  14872   &  14880   &  10526   &  14669   &  14831   & 6.0 &  2.0 &  2.0 \\     
0405680201     &   23503   &  25088   &  25088   &  18915   &  25088   &  25088   & 6.0 &      &      \\
0405750201     &   14907   &  16487   &  16497   &  9107    &  15345   &  15522   & 6.0 &  2.0 &  2.0 \\
0406600301     &   29271   &  29201   &  29471   &  29271   &  29201   &  29471  &&&  \\     
0406600401     &   31885   &  31736   &  32019   &  14822   &  30251   &  31047  & 6.0 &  2.0 &  2.0  \\     
0502170301     &   30462   &  30350   &  30621   &  30462   &  30350   &  30621   &&& \\     
0502370201     &   103307  &  101934  &  101965  &  60587   &  82017   &  84821  & 6.0 &  2.2 &  2.2   \\     
0502370301     &   20452   &  22014   &  22004   &  17455   &  22014   &  22004  & 5.0 &      &        \\     
0503170101     &   54027   &  54422   &  54762   &  53137   &  54380   &  54710  & 6.0 &      &        \\     
0503170201     &   30058   &  31614   &  31619   &  23358   &  31614   &  31619  & 6.0 &      &        \\     
0503170301     &   30487   &  32025   &  32030   &  28987   &  32025   &  32030  & 6.0 &      &        \\     
0510010401     &   10000   &  11945   &  11953   &  6600    &  11685   &  11853   & & & \\     
0511010301     &   7024    &  8583    &  8580    &  1926    &  8281    &  8487    & & & \\
0550451501     &   15963   &  17545   &  17511   &  7565    &  16619   &  17246   & & & \\     
0553950201     &   85608   &  85745   &  85747   &  85608   &  85700   &  85747   & & & \\ 
0554600301     &   37589   &  39302   &  39299   &  31793   &  39302   &  39299  & & &  \\     
0554600401     &   38922   &  40493   &  40503   &  28922   &  34066   &  35163  & & &  \\     
0554720101     &   40863   &  42804   &  42555   &  35866   &  41074   &  41775  & & &  \\     
0555691001     &   5227    &  6815    &  6817    &  5227    &  6815    &  6817   & & &  \\  
0561580101     &   34709   &  34830   &  34825   &  31559   &  34489   &  34746  & & &  \\     
0604090201     &   29033   &  30605   &  30610   &  29033   &  30605   &  30564  & & &  \\     
0651450101     &   0       &  43215   &  43192   &  0       &  42903   &  43168  & & &  \\     
0654190101     &   19999   &  21575   &  21565   &  19602   &  21575   &  21565  & 6.0 &      &       \\     
0654230401     &   37020   &  38603   &  17232   &  27934   &  33938   &  4376   & 6.0 &  2.0 &  2.0  \\
0673930101     &   0       &  23151   &  23127   &  0       &  23151   &  23127  & & &  \\     
0679810401     &   14974   &  15120   &  15123   &  14974   &  15120   &  15123  & & &  \\     
0679810501     &   14477   &  14621   &  14623   &  14477   &  14621  &  14623  & & &  \\     
0690441801     &   83523   &  85124   &  85094   &  75823   &  85124   &  85094  & & &  \\     
0691760101     &   20682   &  22621   &  22626   &  20682   &  22621   &  22626  & & &  \\     
0693900201     &   32942   &  33032   &  33075   &  8014    &  28150   &  29461  & 6.0 &  1.6 &  1.6  \\     
0694030101     &   69714   &  71924   &  71974   &  55161   &  67252   &  69189  & 6.0 &  1.8 &  2.0  \\     
0694040201     &   0       &  43577   &  43584   &  0       &  43577   &  43543   &&& \\
0695000401     &   12041   &  13616   &  13619   &  9741    &  10655   &  11646  & 6.0 &  1.5 &  1.5  \\     
0700980101     &   35703   &  37298   &  37269   &  30808   &  37298   &  37269  &&&  \\
0701230101     &   23891   &  25056   &  25064   &  15450   &  21028   &  21561  & 6.0 &  1.8 &  1.8  \\     
0701230701     &   19682   &  21619   &  21625   &  19283   &  21573   &  21625  & 6.0 &  1.8 &       \\     
0722090101     &   58324   &  59927   &  59913   &  53729   &  59837   &  59913  & 6.0 &  2.0 &       \\     
0722190201     &   128376  &  128427  &  128213  &  102851  &  97845   &  92720  &     &  1.8 &  1.8  \\
0727960901     &   0       &  32236   &  32179   &  0       &  32236   &  32179  & & &  \\
0743980401     &   30040   &  31650   &  31622   &  30040   &  31650   &  31622  & & &  \\     	   
0744600101     &   131717  &  131389  &  135280  &  87598   &  107747  &  116230 & & &  \\     	   
0745130201     &   50799   &  59104   &  59083   &  50600   &  59104   &  59083  & & &  \\     	   
0748391201     &   0       &  40611   &  40589   &  0       &  39491   &  39996  & & &  \\     	   
0762250301     &   110985  &  112591  &  112517  &  81990   &  110010  &  111482 & & &  \\     	   
0763140101     &   89662   &  90905   &  90832   &  37824   &  72710   &  78968  & & &  \\     	   
0763180101     &   14682   &  16648   &  16611   &  2682    &  8335    &  10688  & & &  \\     	   
0763180201     &   32639   &  34239   &  34212   &  27942   &  34239   &  34212  & & &  \\     	   
0763180301     &   30970   &  32579   &  32538   &  30970   &  32584   &  32538  & & &  \\  	   
0782170601     &   11428   &  11610   &  11586   &  7400    &  9773    &  9745   & & &  \\    
0782200101     &   43016   &  44610   &  44572   &  38431   &  42806   &  43155  & & &  \\    
0782770201     &   36903   &  46641   &  46603   &  36903   &  46641   &  46603  & & &  \\    
0783160101     &   101561  &  103193  &  103120  &  81872   &  100636  &  101386 & & &  \\    
0784100401     &   18836   &  20437   &  20403   &  18836   &  20437   &  20403  & & &  \\    
0784860101     &   103763  &  105960  &  106045  &  71989   &  92898   &  95640  & & &  \\    
0790180401     &   34041   &  35411   &  35355   &  26442   &  35411   &  35355  & & &  \\    
0794580301     &   0       &  22167   &  22142   &  0       &  22167   &  22142  & & &  \\    
0795750101     &   0       &  35306   &  35038   &  0       &  35306   &  35038  & & &  \\
0802410101     &   99856   &  101441  &  101428  &  91564   &  101441  &  101428  & 6.0 &      &       \\
0804250301     &   40480   &  40625   &  40605   &  33233   &  40625   &  40605   & & & \\    
0821120101     &   59252   &  61088   &  61099   &  47053   &  60322   &  60579   & & & \\    
0822210101     &   35158   &  37004   &  36986   &  29959   &  37004   &  36986   & 6.0 &      &   \\
0825140101     &   111063  &  112891  &  112894  &  67972   &  109002  &  112031  & 6.0 &      &   \\
0831791301     &   26661   &  25067   &  24994   &  26661   &  25067   &  24994   & & & \\
0840040301     &   21762   &  23620   &  23601   &  15262   &  23516   &  23601   & & & \\     
0840211101     &   26035   &  31058   &  31018   &  12800   &  18342   &  20360   & & & \\     
0844101101     &   23740   &  25602   &  25337   &  16756   &  25450   &  25337   & & & \\     
0845060201     &   25140   &  25621   &  25596   &  23040   &  25621   &  25596   & & & \\
0860140101     &   14761   &  16619   &  16575   &  13761   &  16567   &  16547   & & & \\     
0872392001     &   0       &  54903   &  54158   &  0       &  46061   &  44044   & & & \\
	&&&&\textit{Sgr\,A$^{*}$}&&&  & & \\
0723410301     &   51946   &  53615   &  53605   &  0       &  17073   &  11326   & & & \\     
0723410401     &   54023   &  55618   &  55586   &  42428   &  55484   &  55586   & & & \\     
0723410501     &   54886   &  59792   &  59778   &  38852   &  49943   &  50080   & & & \\     
0724210201     &   55585   &  57165   &  57170   &  41687   &  56177   &  56650   & & & \\
0724210501     &   39374   &  40970   &  40932   &  29533   &  39637   &  40584   & & & \\
0743630201     &   32039   &  33642   &  33619   &  17239   &  32871   &  33203   & & & \\     
    0743630301     &   25039   &  26641   &  26619   &  25039   &  26641   &  26619  & & &  \\     
0743630401     &   25740   &  32754   &  32815   &  16244   &  27118   &  28200  & & &  \\     
0743630501     &   37841   &  39447   &  39419   &  37341   &  39101   &  39120  & & &  \\     
0743630601     &   30433   &  32030   &  31994   &  7200    &  27265   &  29784  & & &  \\     
0743630801     &   24030   &  25631   &  25590   &  20634   &  25631   &  25590  & & &  \\     
0743630901     &   28094   &  28008   &  27875   &  15767   &  25700   &  26541  & & &  \\
0822680201     &   26460   &  28315   &  28299   &  26360   &  28315   &  28299  & & &  \\    
0822680301     &   32227   &  36059   &  36039   &  6500    &  24613   &  26364  & & &  \\    
0822680401     &   32260   &  34104   &  34100   &  30560   &  34104  &  34100  & & &  \\    
0822680501     &   29053   &  36313   &  36268   &  21200   &  24957   &  25508  & & &  \\    
0831800101     &   8762    &  10614   &  10600   &  8762    &  10614   &  10548  & & &  \\
0831800201     &   11762   &  13614   &  13598   &  11662   &  13614   &  13598  & & &  \\
0831800301     &   10762   &  12615   &  12600   &  10762   &  12615   &  12600  & & &  \\
0831800401     &   10761   &  12615   &  12600   &  10761   &  12615   &  12600  & & &  \\
0831800501     &   11762   &  13614   &  13601   &  11762   &  13510   &  13497  & & &  \\
0831800601     &   11761   &  13614   &  13598   &  6761    &  13614   &  13598  & 7.5 &      &      \\
0831800701     &   11462   &  13314   &  13299   &  11462   &  13314   &  13299  & & &  \\
0831800801     &   10761   &  12615   &  12600   &  10761   &  12615   &  12600  & & &  \\
0831800901     &   12862   &  14715   &  14698   &  4362    &  14611   &  14698   & 7.0 &      &      \\
0831801001     &   18762   &  20615   &  20600   &  18762   &  20355   &  20392   &     &  2.0 &  2.0 \\
0831801101     &   14762   &  16616   &  16599   &  2562    &  13600   &  16443   & 7.0 &  2.0 &      \\
0831801201     &   20729   &  22580   &  22568   &  5062    &  13516   &  14332   & 7.0 &  2.2 &  2.2 \\
0831801301     &   14196   &  16041   &  16037   &  11200   &  15477   &  16037   & 2.0 &      &      \\
0831801401     &   7929    &  9777    &  9774    &  7929    &  9777    &  9774  &     &  2.2 &  2.2    \\
0831801501     &   7262    &  9115    &  9098    &  7262    &  9115    &  9098   & & &  \\
0831801601     &   10259   &  12099   &  12088   &  5163    &  12099   &  12088  & & &  \\
0851180901     &   68214   &  70063   &  70029   &  53531   &  63018   &  63464  & & &  \\
	&&&&\textit{GCL}&&	 & & & \\
0764190101     &   34845   &  36464   &  36403   &  27627   &  34338   &  34893  & & &  \\     
0764190201     &   25041   &  26649   &  26619   &  18142   &  26337   &  26411  & & &  \\     
0764190301     &   30040   &  31650   &  31619   &  27741   &  31650   &  31619  & & &  \\     
0764190401     &   30736   &  32349   &  32303   &  17136   &  32037   &  32269  & & &  \\     
0764190501     &   30335   &  31950   &  31908   &  6341    &  30286   &  31297  & & &  \\     
0764190601     &   25036   &  26614   &  26619   &  14141   &  24517   &  25111  & & &  \\     
0764190701     &   25033   &  26641   &  26604   &  3839    &  24624   &  25200  & & &  \\     
0764190801     &   33378   &  38327   &  38301   &  4600    &  31285   &  31616  & & &  \\     
0764190901     &   25009   &  26612   &  26593   &  21809   &  25317   &  25301  & & &  \\     
0764191001     &   30024   &  31634   &  31601   &  27225   &  31634   &  31601  & & &  \\     
0764191101     &   30036   &  31642   &  31598   &  22136   &  30966   &  31515  & & &  \\     
0764191201     &   30041   &  31650   &  31619   &  29341   &  31650   &  31619  & & &  \\     
0764191301     &   36678   &  38341   &  38319   &  33637   &  37905   &  37871  & & &  \\     
0764191401     &   27541   &  29150   &  29120   &  24941   &  29098   &  29120  & & &  \\     
0764191501     &   14467   &  29144   &  29118   &  14467   &  29144   &  29118  & & &  \\     
0764191601     &   16041   &  17647   &  17620   &  15841   &  17647   &  17620  & & &  \\     
0764191701     &   32934   &  34534   &  34503   &  32341   &  34534   &  34503  & & &  \\     
0764191801     &   25041   &  26639   &  26619   &  24541   &  26639   &  26619  & & &  \\
0801680101    &   25036   &  26647   &  26604   &  25036   &  26647   &  26604   & & & \\     
0801680201    &   25033   &  26641   &  26619   &  19135   &  24777   &  25475   & & & \\     
0801680301    &   25040   &  26649   &  26619   &  9541    &  16301   &  17467   & & & \\     
0801680401    &   25040   &  26649   &  26619   &  19841   &  24829   &  25319   & & & \\     
0801680501    &   40599   &  41672   &  41622   &  22497   &  27995   &  28400   & & & \\
0801680601    &   27041   &  28650   &  28621   &  17341   &  27093   &  27930   & & & \\     
0801680701    &   26974   &  28577   &  28543   &  26578   &  28577   &  28543   & & & \\     
0801680801    &   26937   &  28537   &  28520   &  25538   &  28444   &  28416   & & & \\     
0801681201    &   34006   &  35019   &  34984   &  28707   &  25433   &  25392   & & & \\     
0801681301    &   25041   &  26649   &  26619   &  25041   &  26649   &  26619   & & & \\
0801681401    &   25041   &  26649   &  26619   &  24641   &  26441   &  26463   & & & \\     
0801681501    &   30627   &  32217   &  32194   &  15629   &  28180   &  28681   & & & \\     
0801681601    &   34001   &  35022   &  34932   &  32009   &  25521   &  25491   & & & \\  
0801681701    &   25038   &  26641   &  26609   &  25038   &  26641   &  26609  & & &  \\     
0801681801    &   25038   &  26644   &  26609   &  25038   &  26644   &  26609  & & &  \\     
0801681901    &   24696   &  26528   &  26523   &  12799   &  26084   &  26523  & & &  \\     
0801682001    &   26748   &  28593   &  28598   &  26748   &  28593   &  28546  & & &  \\     
0801682101    &   28852   &  30321   &  30348   &  19875   &  28776   &  29319  & & &  \\     
0801682201    &   39039   &  40638   &  40589   &  21839   &  39713   &  40516  & & &  \\     
0801682301    &   26986   &  28823   &  28824   &  26888   &  28823   &  28824  & & &  \\     
0801682401    &   29468   &  31523   &  31832   &  17414   &  0&  25309   &	         & &  \\     
0801682501    &   0       &  27613   &  27599   &  0       &  27613   &  27599  & & &  \\     
0801682601    &   21032   &  27616   &  27599   &  20832   &  27616   &  27599  & & &  \\     
0801682801    &   29642   &  31477   &  31486   &  29344   &  31477   &  31486  & & &  \\     
0801682901    &   24711   &  26555   &  26557   &  24717   &  26555   &  26557  & & &  \\     
0801683001    &   26437   &  28024   &  28013   &  11276   &  21459   &  24329  & & &  \\     
0801683101    &   30562   &  32403   &  32399   &  30562   &  32403   &  32399  & & &  \\     
0801683201    &   25748   &  27508   &  27493   &  22970   &  25812   &  25835  & & &  \\     
0801683301    &   26532   &  28126   &  28104   &  24936   &  28126   &  28104  & & &  \\     
0801683401    &   26040   &  27648   &  27620   &  26040   &  27648   &  27620  & & &  \\     
0801683501    &   28259   &  30103   &  30085   &  27559   &  30053   &  29942  & & &  \\     
0801683601    &   28428   &  30278   &  30257   &  21029   &  29186   &  29906  & & &  \\     
0824930701     &   28392   &  32401   &  32355   &  13632   &  24427   &  25759 & & &   \\     
0824930801     &   24762   &  26614   &  26598   &  24762   &  26614   &  26598 & & &   \\     
0824930901     &   24504   &  26350   &  26333   &  24504   &  26350   &  26333 & & &   \\     
	&&&&\textit{CMZ}&&	& & &  \\
0862470101     &   47435   &  49273   &  49260   &  46639   &  49273   &  49260 & & &   \\     
0862470201     &   39739   &  41581   &  41559   &  36139   &  41425   &  41486 & & &   \\     
0862470301     &   43374   &  45239   &  45204   &  41780   &  45239   &  45181 & & &   \\     
0862470401     &   46234   &  48185   &  48174   &  40850   &  46896   &  47513 & & &   \\     
0862470501     &   45758   &  47610   &  47586   &  22900   &  47202   &  47391 & & &   \\     
0862470601     &   39759   &  41602   &  41601   &  39559   &  41602   &  41601 & & &   \\     
0862470701     &   39694   &  41540   &  41515   &  35096   &  40350   &  40449 & & &   \\     
0862470801     &   39740   &  41581   &  41554   &  32246   &  39889   &  40766 & & &   \\     
0862470901     &   43951   &  45863   &  45839   &  19611   &  31724   &  32953 & & &   \\     
0862471001     &   37571   &  41162   &  41112   &  29159   &  30585   &  29609 & & &   \\     
0862471101     &   40739   &  42596   &  42586   &  29048   &  34648   &  35319 & & &   \\     
0862471201     &   42969   &  45520   &  45498   &  34400   &  38997   &  40194 & & &   \\     
	&&&&\textit{Heritage}&& &&&\\
0886010101     &   29932   &  33403   &  33275   &  22504   &  28381   &  28575  & 6.0 &  2.0 &  2.0  \\     
0886010201     &   19759   &  21619   &  21599   &  14462   &  20475   &  20663  & 7.0 &  2.0 &  2.0  \\     
0886010301     &   18907   &  20739   &  20738   &  9562    &  19259   &  20696  & 7.0 &  2.0 &       \\     
0886010401     &   19762   &  21619   &  21599   &  18862   &  21619   &  21599  & 6.0 &      &       \\     
0886010501     &   19742   &  21600   &  21557   &  15550   &  21600   &  21557  & 6.0 &      &       \\     
0886010601     &   19755   &  21619   &  21599   &  19755   &  21619   &  21599   &&& \\
0886010701     &   19762   &  21619   &  21599   &  18162   &  21619   &  21599   & 6.0 &      &      \\     
0886010801     &   19760   &  21619   &  21599   &  15260   &  21619   &  21599    & 6.0 &      &     \\
0886010901     &   19761   &  21619   &  21599   &  19761   &  21619   &  21599  & & &  \\     
0886011001     &   19762   &  21619   &  21599   &  19762   &  21619   &  21599  & & &  \\     
0886011101     &   22720   &  24578   &  24560   &  22720   &  24578   &  24560  & & &  \\     
0886011201     &   22740   &  24595   &  24576   &  15740   &  24595   &  24576  & 7.0 &      &   \\     
0886011301     &   23182   &  25027   &  24981   &  23182   &  25027   &  24981  & & &  \\     
0886020101     &   19678   &  21530   &  21500   &  19678   &  21530   &  21500  & & &  \\     
0886020201     &   19752   &  21568   &  21541   &  19752   &  21568   &  21541  & & &  \\     
0886020301     &   19754   &  21619   &  21599   &  19754   &  21619   &  21599  & & &  \\     
0886020401     &   23192   &  25024   &  25018   &  9145    &  24107   &  24505 & 7.0 &      &    \\     
0886020501     &   19754   &  21614   &  21591   &  11158   &  21614   &  21547  & 7.0 &      &   \\     
0886020701     &   32932   &  35003   &  34971   &  24800   &  28545   &  31185  & 7.0 &    2.3  & 2.3   \\     
0886020801     &   21907   &  23763   &  23749   &  21807   &  23763   &  23749  & 6.0 &      &    \\     
0886020901     &   20501   &  22346   &  22342   &  19301   &  22346   &  22342   & 7.0 &      &  \\     
0886021201     &   19691   &  21549   &  21528   &  19691   &  21549   &  21528  & & &  \\     
0886021301     &   29850   &  31701   &  31684   &  29850   &  31701   &  31684  & & &  \\     
0886030101     &   19762   &  21619   &  21599   &  19762   &  21619   &  21599  & & &  \\     
0886030201 & 25101 &  32498 & 32258 & 17836 &  20745 &  20777  & & & \\
 0886030301 & 19729 &  21584 & 21562 & 19729 &  21535 &  21562  & & & \\
 0886030401 & 21758 &  23609 & 23598 & 21758 &  23609 &  23598  & & & \\
 0886030501 & 19726 &  21568 & 21560 & 19726 &  21568 &  21560  & & & \\
 0886030601 & 19750 &  21595 & 21565 & 19750 &  21595 &  21565  & & & \\
 0886030701 & 21626 &  25094 & 25059 & 7500  & 18457  & 19684   & & & \\
 0886030801 & 19449 &  21619 & 21599 & 18749 &  21619 &  21599  & & & \\
 0886030901 & 19756 &  21598 & 21599 & 17856 &  21598 &  21599  & & & \\
 0886031001 & 22381 &  27986 & 26705 & 12746 &  17729 &  16769  & & & \\
 0886041101     &   19762   &  21619   &  21599   &  16062   &  21619   &  21599  & & &  \\  
 0886060101 & 23959 &  25810 & 25800 & 23959 &  25810 &  25800  & & & \\
 0886060201 & 24762 &  26609 & 26601 & 20362 &  26609 &  26601  & & & \\
 0886060301 & 19761 &  21619 & 21599 & 19262 &  21619 &  21599  & & & \\
 0886060501 & 22353 &  26420 & 26138 & 8598  & 17513  & 19475   & & & \\
 0886060601 & 24749 &  26595 & 299   & 16753 &  22706 &  271    & & & \\
 0886060701 & 21456 &  23296 & 15561 & 21357 &  23255 &  15561  & & & \\
 0886060801 & 19761 &  21619 & 21599 & 19761 &  21619 &  21547  & & & \\
 0886060901 & 21559 &  23420 & 23387 & 21559 &  23420 &  23387  & & & \\
 0886061001 & 19758 &  21622 & 21588 & 19059 &  21622 &  21588  & & & \\
 0886061101 & 42809 &  46600 & 46714 & 29985 &  36337 &  37208  & & & \\
 0886070101 & 19707 &  21563 & 21521 & 13807 &  17149 &  16979  & & & \\
 0886070201 & 23932 &  27120 & 27589 & 2896  & 11804  & 13625   & & & \\
\end{longtable}
\clearpage
\twocolumn

\end{appendix}


\end{document}